\documentclass[]{article}
\usepackage{fullpage}
\usepackage{amsmath}
\usepackage{graphicx}
\usepackage[round]{natbib}  
\usepackage[colorlinks=true,citecolor=blue,linkcolor=black,urlcolor=blue]{hyperref}
\usepackage{multirow}
\usepackage{float,epsf,caption,subcaption}
\usepackage{longtable}
\usepackage{tabularx}
\usepackage{authblk}
\usepackage{setspace} 

\usepackage[english]{babel}
\usepackage{pdflscape} 

\title{Real-time large-scale supplier order assignments across two-tiers of a supply~chain with penalty and dual-sourcing}

\author[1,2]{Vinod Kumar Chauhan\thanks{vinod.kumar@eng.ox.ac.uk (This work was done at University of Cambridge UK.)}}
\author[1]{Stephen Mak\thanks{sm2410@cam.ac.uk}}
\author[1]{Ajith Kumar Parlikad\thanks{aknp2@cam.ac.uk}}
\author[3]{Muhannad Alomari\thanks{Muhannad.Alomari@Rolls-Royce.com}}
\author[3]{Linus Casassa\thanks{Linus.Casassa@Rolls-Royce.com}}
\author[1]{Alexandra Brintrup\thanks{ab702@cam.ac.uk (corresponding author)}}
\affil[1]{Institute for Manufacturing, Department of Engineering, University of Cambridge UK}
\affil[2]{Institute of Biomedical Engineering, Department of Engineering Science, University of Oxford UK}
\affil[3]{R$^2$ Data Labs, Rolls-Royce}

\begin{document}
	
	\maketitle
	
	\begin{abstract}
		Supplier selection and order allocation (SSOA) are key strategic decisions in supply chain management which greatly impact the performance of the supply chain. Although, the SSOA problem has been studied extensively but less attention paid to scalability presents a significant gap preventing adoption of SSOA algorithms by industrial practitioners.
		This paper presents a novel multi-item, multi-supplier double order allocations with dual-sourcing and penalty constraints across two-tiers of a supply chain, resulting in cooperation and in facilitating supplier preferences to work with other suppliers through bidding.
		We propose Mixed-Integer Programming models for allocations at individual-tiers as well as an integrated allocations.
		An application to a real-time large-scale case study of a manufacturing company is presented, which is the largest scale studied in terms of supply chain size and number of variables so far in literature. The use case allows us to highlight how problem formulation and implementation can help reduce computational complexity using Mathematical Programming (MP) and Genetic Algorithm (GA) approaches. The results show an interesting observation that MP outperforms GA to solve SSOA.
		Sensitivity analysis is presented for sourcing strategy, penalty threshold and penalty factor. The developed model was successfully deployed in a large international sourcing conference with multiple bidding rounds, which helped in more than 10\% procurement cost reductions to the manufacturing company.

	\end{abstract}

	\section{Introduction}
	\label{sec_intro}
	
	Supplier selection and order allocation (SSOA) is the procurement problem of determining which item/material should be procured from which suppliers in what quantities, that has been widely studied (\cite{Chakravarty1979,Pan1989}). SSOA constitutes a fundamental part of supply chain management which greatly impact the performance and competitiveness of the supply chain (\cite{Amid2006}). In many industries, the procurement of materials and services can cost up to 80\% of the total cost of a product (\cite{Willard2012}). 
	
	Having been studied for few decades, research streams on SSOA have focused on the creation of optimization algorithms and mathematical models for various problem types, including single, dual, multi-sourcing, the incorporation of discounting and inflation, and how closed-loop supply chain and sustainable configurations could be embedded (\cite{Aouadni2019,Pasquale2020,naqvi2021supplier}).
	
	However, according to a recent review, only 34\% of the studies in the sample were conducted based on real industrial cases or real collected data (\cite{Pasquale2020}). The literature consists mainly of supply chains of limited sizes - our review shows that for real cased studies, the maximum scale of supply chain studied under the SSOA literature consists of 15 suppliers and 90,000 variables (See Table~\ref{tab_scale}). 
	
	In light of the above, the industrial trend shows an interesting conundrum. Companies with large scale complex supply chains would be more likely to be in need of automated algorithms to configure their supply chains, as the problem space becomes too complex to tackle manually. Recent studies show an increasing trend in the complexity of supply chains (e.g. \cite{Christopher2021}), add to the urgency of need for algorithms that can handle high numbers of products and suppliers in SSOA optimization.
	
	Large-scale SSOA across multiple-tiers brings about unique advantages as well as challenges. Current little to no attention paid to the multi-tier and large-scale nature of the problem presents a gap between theory and practice, as extant algorithms are not adopted to serve the needs of industries with complex supply networks including aerospace engineering, industrial machinery and medical devices. Taking a multi-tier perspective, where suppliers’ preferences to work with one another are incorporated is realistic and helps improve cooperation. Large-scales of complex supply networks are another reality that have not yet been tackled. Several industries have to select and assign hundreds, if not thousands of products/items during procurement due to the complexity of engineered products being built. However, as our literature review shows, scalability of existing approaches are hitherto unknown. On the other hand, the solution approach we present that tackles these issues do increase the complexity of the problem due to the inherent dependency of the tiers on each other. Moreover, the collection of data for multiple tiers is difficult and the OEM needs to be in a position to be able to do so, for our approach to be applicable.
	
	In this paper, we present a set of techniques for researchers to consider solving the SSOA at scale and discuss how it can be extended to SSOA variants. We test our proposed approach using a real-life case study involving 2 tiers, 7,200,020 decision variables and 70 suppliers, constituting the largest SSOA use case to date. The case study includes a time constraint, in that the SSOA allocation needs to be done under a reasonable time limit that would allow procurement officers to negotiate with suppliers proposing bids during a multi-round bidding process taking place during a sourcing conference. Our work presents a novel SSOA problem which considers allocations on two-tiers allowing suppliers to bid to work with each other and penalize non-preferred allocations through the penalty constraints. The problem characteristics additionally include NP-hardness and dual-sourcing constraints.
	
	Our results show that there exists few methods in literature that can handle a real-time constraint. Problem formulation, as well as modelling language and solver engine all play a role in the success of the algorithm, whereas extant literature typically treats these solution components in isolation from one another. We go on to argue that all three components need to be considered as they impact one another.
	
	Thus, the research question of the study is given as: How to automate a dynamic/real-time real-life large-scale double supplier allocations at two-tiers of a multi-item supply chain with dual-sourcing and penalty constraints? The contributions of the study, while solving the research questions are summarized below.
	\begin{itemize}
	    \item A novel multi-item multi-supplier order allocation problem with penalty constraints and dual-sourcing is presented.
        \item The problem involves double allocations at two different tiers of the supply chain which result in cooperation between tiers and facilitate supplier preferences to work with each other.
        \item Mixed-Integer Linear Programming models are developed for supplier order allocations at individual-tiers as well as  integrated allocation.
        \item A generic approach to solve large-scale problems using mathematical programming is presented.
        \item A dynamic large-scale real-life case study of a manufacturing is presented which helps the company to automate the manual allocations and negotiate better prices.
	\end{itemize}
	
	The rest of the paper is organized as follows: 
	
	Section~\ref{sec_literature} gives the relevant background of the SSOA problem highlighting a significant gap on real-life studies that consider problem scale. Section~\ref{sec_problem} introduces the case study which sets out the industrial context in which large scale SSOA issues arise. Section~\ref{sec_solution} proposes a threefold method to handle large scale SSOA problems. Section~\ref{sec_experiments} applies the aforementioned method to the case study presenting experimental results. Section~\ref{sec_conclusion} concludes the paper.

	\section{Literature review}
	\label{sec_literature}
	Here, a brief literature review on supplier selection and order allocation (SSOA), discounts on order allocation and scale of the problem are presented.
	
	\subsection{SSOA}
	\label{sec_overview}
	SSOA problem has attracted the interests of academicians as well as practitioners for a long time with the earliest known works starting in 1979 (\cite{Chakravarty1979}). Here, we briefly summarize the main categorisations of the SSOA problem to give the reader a background, before moving onto discounting and industrial practicability, which form the main constituencies of our inquiry. For detailed reviews on the SSOA problem, please refer to \cite{Aouadni2019,Pasquale2020}, and \cite{naqvi2021supplier}.
	
	The problem has been studied extensively in different industrial settings including automotive (\cite{Moheb-Alizadeh2019}), manufacturing (\cite{Guo2013}) and food supply chain (\cite{Mohammed2018}) etc., and with different supply chain configurations, e.g., traditional SC (\cite{Chakravarty1979}), green/sustainable SC (\cite{HOSSEINI2022107811}), collaborative/Integrated SC (\cite{Renna2015}) and closed-loop supply chains SC (\cite{nasr2021novel}).
	
	Different settings included supply chain configuration under disruption risks (\cite{PrasannaVenkatesan2016}), supply chains with single and multiple products, single, dual and multi-sourcing strategies (\cite{Sawik2014b}), with discounts (\cite{Alegoz2019}), inflation (\cite{Khoshfetrat2020}) and order splitting (\cite{SUN2022105515}) etc.
	
	SSOA can also be classified into three categories, according to the solution approach (\cite{Pasquale2020}): configuration problems involve those where (i) Suppliers are selected from a predefined certified list, according to some measure such as reliability (\cite{Meena2013}) (ii) Bi-phase SSOA where suppliers are selected using a multi-criteria decision making (MCDM) method such as analytic hierarchy process (AHP), analytical network process (ANP) and technique for order preference by similarity to ideal solution (TOPSIS) or artificial intelligence (AI) based techniques (\cite{Ccebi2016}), after which orders are allocated. (iii) An integrated model is used for supplier selection and order allocation (\cite{Gupta2016}).
	
	SSOA problems have been solved mainly by four types of methods: (i) Mathematical Programming methods such as linear, non-linear integer programming, single and multi-objective optimization problems etc. (\cite{Sawik2014a}), 
	(ii) MCDM methods such as AHP, ANP, VIKOR (which stands for `VlseKriterijumska Optimizacija I Kompromisno Resenje') and TOPSIS etc. (\cite{Alegoz2019}), 
	(iii) AI methods such as neural networks, fuzzy inferencing, GAs and particle swarm optimization (PSO) etc. (\cite{Meena2013}), and 
	(iv) Simulation based methods (\cite{Moghaddam2015}).

	\subsection{Penalties and discounts}
	\label{sec_discounts}
	Discounts in SSOA have been studied by many researchers, such as \cite{Alegoz2019,Alfares2018,Ayhan2015,Cheraghalipour2018,Hadian2018,Meena2013,Moheb-Alizadeh2019,Safaeian2019,Shalke2018}.
	
	Some notable works include:
	
	\cite{VitalSoto2017} addressed multi-period lot sizing problem with supplier selection and considered both all-units and incremental quantity discounts. A mixed-integer non-linear programming (MINLP) model is developed and solved using a hybridized search evolutionary LP-driven local search method. 
	
	\cite{Ghaniabadi2017} considered dynamic lot sizing with supplier selection, backlogging and both all-units and incremental quantity discounts. They proposed mixed-integer linear programming (MILP) models and solved using Gurobi 6.5.2.
	
	\cite{Shalke2018} studied all-unit and incremental quantity discounts in a sustainable supply chain. They proposed a multi-objective model for SSOA and solved it using a multi-choice goal programming approach.
	
	\cite{Cheraghalipour2018} studied a case study of SSOA in plastic industry with disruption risks and considered quantity discounts of all-unit, incremental and no-discounts. Here, supplier selection is performed using a multiple criteria decision making (MCDM) approach called best-worst method (BWM) and for order allocation a MILP model is proposed to minimize the total costs and maximize the total sustainability score of all suppliers. The proposed model is solved using a Revised Multi-Choice Goal Programming (RMCGP) method. 
	
	Most recently, \cite{Alegoz2019} posed an SSOA problem with fast service options and considered incremental quantity discount and no-discounts. They developed a hybrid approach based on fuzzy TOPSIS (FTOPSIS), trapezoidal type-2 fuzzy AHP and goal programming.
	
	A challenge in multi-tier SSOA that has not yet been handled is price dependencies resulting from relations between suppliers. When a certain Tier1 supplier is selected to supply a certain product, that supplier itself will have preferences on whom it wants to work with on Tier2. However, if Tier2 suppliers are also suggested by the OEM, such as in the use case presented here, then the Tier1 supplier will assert its preferences by applying different bids. Another interesting point to note is that when a certain amount of orders is given to a Tier1 supplier, unit price of obtaining parts from Tier2 may increase or decrease, which will be passed onto the OEM. While most SSOA problems to date either assumed no control by the OEM over the second tier, or full control, where the suppliers choices to work with one another are not considered. Our study presents an approach where such dependencies are considered through the incorporation of penalties from suppliers. The penalties are opposite of discounts provided by suppliers and not discussed for SSOA. Moreover, these penalty constraints are based on certain type of items and orders received by suppliers, unlike quantity discounts studied in the literature.

	\subsection{Scale and industrial practicability}
	\label{sec_recent}
	Although the SSOA problem has been studied extensively in different settings, according to a recent systematic review by \cite{Pasquale2020}, only 34\% of the studies in the sample were conducted based on real industrial cases or real collected data. Even in recent studies, supply chains of limited sizes were studied. Typically, the studies consider 5--10 suppliers and 5--20 products/parts, e.g., automotive studies (see Table~\ref{tab_scale}). But in reality, these industries have thousands of parts and large number of suppliers.
	
	Some recent example studies are summarized here. 
	
	\cite{Sawik2014a} considered a customer driven supply chain where different parts supplied by different suppliers are assembled into a variety of products by a producer to meet customer orders. The problem considered a numerical study with 10 suppliers and 25 products, each of which could have up to 3 parts. Both single and dual-sourcing under disruption risks were considered. The authors formulated a mixed-integer programming model and solved using AMPL (A Mathematical Programming Language) modelling language and CPLEX solver. This work was further extended using multiple sourcing in \cite{Sawik2014b} which was solved using CPLEX and Gurobi solvers.
	
	\cite{Moheb-Alizadeh2019} solved Sustainable SSOA problem for multiple products, multiple periods in a multimodal transportation supply chain with shortage and discount conditions. They formulated a multi-objective MILP (MOMILP) model which was solved using a hybrid solution approach based on Benders decomposition. They studied a use case from the automotive industry with 4 suppliers and 2 raw materials for 2 years with 172 variables. But they studied a large test problem with 50 suppliers, 30 materials and 18 products with 6,929,314 variables. This is the largest study to date with respect to number of variables.
	
	\cite{Yousefi2019} considered a two-stage hybrid supply chain with single-buyer and multi-vendor coordination. They used a novel, game theoretic, pricing strategy. The problem was formulated as multi-objective MINLP (MOMINLP) model to minimize costs and evaluate suppliers simultaneously, which was solved using LINGO 14 software. It was a numerical study and the problem had 1 buyer and 10 suppliers with less than 100 variables.
	
	\cite{Alegoz2019} considered fast service options and discount factors in SSOA problem and developed a hybrid approach based on FTOPSIS, trapezoidal type-2 fuzzy AHP and goal programming. They considered a case study of 6 suppliers and 13 products with 325 variables.
	
	\cite{Hosseini2019} considered SSOA problem for building resilient SC considering supplier restoration under disruption risks. They used a probabilistic graphical model for supplier selection and stochastic multi-objective model for order allocation is solved using a fuzzy c-mean clustering algorithm and using augmented $\epsilon$-constraint method. They studied numerical problems with 10-19 suppliers and 397-861 variables. 
	
	\cite{Mohammed2019} studied the sustainable supply chain of a metal factory with 3 suppliers in Saudi Arabia, and having only 3 variables. For supplier evaluation and ranking, they considered economic, environmental and social aspects using an integrated Fuzzy AHP (FAHP)--FTOPSIS method, and for selecting supplier and assigning optimal quantities, using a fuzzy multi-objective optimization model. 
	
	\cite{EsmaeiliNajafabadi2019} considered a centralized supply chain with disruption risks and formulated the SSOA problem as a MINLP model. Their use case had 1 buyer, 10 suppliers and 2 types of parts. 
	
	\cite{Almasi2019} studied a sustainable supply chain of an automotive manufacturing company under disruption risk and inflation. They formulated the SSOA problem as a multi-objective and multi-period mathematical model, and solved using weighted sum approach (WSA) and augmented $\epsilon$-constraint (AEC) method. The supply chain had 1 manufacturer, 5 products, 3 suppliers and 3 periods, and 750 variables in the model.
	
	\cite{Rezaei2020} discussed SSOA problem in closed-loop supply chain configuration with various sourcing strategies and under disruption risk. They used the sample average approximation (SAA) method to solve the problem which considered a numerical example having 30 suppliers, 20 products and 5 parts with 78,152 variables.
	
	{\scriptsize
		\begin{landscape}
		\begin{longtable}[htb!]{p{0.15\linewidth}p{.10\linewidth}p{.01\linewidth}p{.10\linewidth}p{.10\linewidth}p{.05\linewidth}p{.35\linewidth}}
			\caption{Recent studies on SSOA problem and their scale (\#S = suppliers, \#P = number of products/parts, \#V = number of variables).}
			\label{tab_scale}\\
			\hline
			\textbf{References} & \textbf{Study} & \textbf{\#S} & \textbf{\#P} & \textbf{Others} & \textbf{\#V} & \textbf{Techniques} \\ \hline \endhead
			
			\cite{Wu2022} & numerical & 18 & -- & 50 orders, 7 vehicles & $<$144,000 & An improved shuffled frog-leaping algorithm solved a hybrid cross-supplier order assignments and third-party logistics scheduling model.\\ 
			
			\cite{SUN2022105515} & numerical & 6 & 1 & 20 retailers & -- & An MINLP solved analytically using MATLAB and verified through simulation.\\ 
			
			\cite{HOSSEINI2022107811} & composite products & 5 & 3 & 3 items, 2 periods & -- & BWM-evidential reasoning for supplier evaluation and bi-objective allocation solved using stochastic and dynamic programming.\\ 
			
			\multirow{2}{*}{\cite{Vishnu2021}} & footwear manuf. & 8 & -- & 3 plants, 12 distribution centres, 9 customer zones & 567 & Multi-objective model with FTOPSIS used for assigning weights of different objectives and GA used to solve the model.\\ 
			& numerical & 200 & -- & 50 plants, 50 distribution centres, 220 customer zones & 71,240 & \\ 
			
			\cite{Mohammed2021} & manuf. & 5 & -- & -- & 5 & AHP and TOPSIS used for supplier evaluation, $\epsilon$-constraint method solved multi-objective optimization problem and TOPSIS selected the final Pareto solution.\\ 
			
			\cite{Li2021} & empirical & 10 & 4 products & 3 period & 240 & BWM used for primary supplier evaluation and MATLAB used to solve multi-objective model for dynamic supplier selection and order allocation.\\ 

			\cite{nasr2021novel} & garment industry & 5 & 3 prod, 3 sub-prod & 2 raw materials, 3, 4, 3, 3, 3 production, distribution, collection, reproduction, disposal centers, 10 customers, 3 price levels, 8 vehicles, 2 transport modes, 6 periods & 21,472 & fuzzy-BWM for supplier evaluation and MOMILP for order allocation solved using GAMS.\\ 
			
			\cite{RezaeiA2020} & car manuf. & 3 & -- & -- & 6 & AHP and FAHP used for supplier evaluation and MATLAB used to solve bi-objective model for order allocation.\\ 
			
			\cite{Khoshfetrat2020} & auto. ind. & 3 & 5 products & 3 period & - & AHP used for supplier evaluation and LINGO 11 software used to solve multi-objective order allocation model.\\ 
			
			\cite{Rezaei2020} & numerical & 30 & 5 parts, 20 products & -- & 78,152 & SAA method used to solve the order allocation.\\ 
			
			\cite{Feng2020} & auto. manuf. & 5 & 3 products & 5 periods & 145 & Linguistic entropy weight method used for supplier evaluation and multi-objective order allocation model solved using Lingo 11 software.\\ 
			
			\cite{Shaikh2020} & numerical & 3 & 2 items & -- & 6 & A multi-objective order allocation model was solved using GA.\\ 
			
			\cite{Bektur2020} & medical devices & 6 & 3 products & 3 periods & 242 & FAHP and fuzzy Preference Ranking Organization Method for Enrichment Evaluation (F-PROMETHEE) used for supplier evaluation and AEC method and LP-metrics for solving multi-objective allocation model to get Pareto solution. Final solution was selected using TOPSIS.\\
			
			\cite{Yaghin2020} & empirical & 6 & 2 products & 2 transport modes, & 96 & An integrated multi-objective model for supplier selection, order allocation and transport planning was solved using a novel flexible-possibilistic chance constraint approach.\\ 
			
			\cite{Safaeian2019} & numerical & 9 & -- & 3-8 discount levels & 99 & The Zimmermann fuzzy approach was used to convert multi-objective model to single objective and GA was used to solve the problem.\\ 
			
			\cite{Yan2019} & numerical & 2 & -- & & 2 & An analytical closed-form solution was developed for order allocation.\\ 
			
			\multirow{2}{*}{\cite{Moheb-Alizadeh2019}} & auto. ind. & 4 & 2 materials, 3 products & 2 period, 3 transport modes, 1 purchasing firm, 2 discount levels & 172 & A multi-objective model was solved using a hybrid solution approach based on Benders decomposition.\\ 
			& numerical & 50 & 30 materials, 18 products & 24 periods, 10 transport modes, 8 purchasing firms, 2 discount levels & 6,929,314 & \\ 
			
			\cite{Yousefi2019} & numerical & 10 & -- & & $<$100 & A multi-objective model was solved using LINGO 14 software.\\ 
			
			\cite{Alegoz2019} & empirical & 6 & 13 products & -- & 325 & A hybrid approach based on FTOPSIS, trapezoidal type-2 FAHP and goal programming was developed for SSOA.\\
			
			\cite{Hosseini2019} & numerical & 19 & -- & & 861 & A probabilistic graphical model was used for supplier evaluation and stochastic multi-objective model for order allocation was solved using a fuzzy c-mean clustering algorithm and using AEC method.\\
			
			\cite{Mohammed2019} & metal factory & 3 & -- & & 3 & An integrated FAHP-FTOPSIS method was used for supplier evaluation and a fuzzy multi-objective optimization model was used for SSOA.\\
			
			\cite{EsmaeiliNajafabadi2019} & numerical & 10 & 2 parts & -- & & An MINLP model solved using General Algebraic Modeling System (GAMS) 24.8.\\ 
			
			\cite{Almasi2019} & auto. manuf. & 5 & 5 products & 3 periods & 750 & A multi-objective model was solved using WSA and AEC method.\\ 
			
			\cite{Yaghin2019} & numerical & 5 & 3 items & 4 periods & 200 & A fuzzy multi-objective MINLP model solved using fuzzy multi-choice goal programming.\\ 
			
			\cite{Sontake2019} & auto. & 5 & 3 parts & -- & 60 & AHP and TOPSIS were used for supplier selection and the knapsack problem was used for order allocation.\\ 
			
			\cite{Mari2019} & garments & 10 & -- & & 110 & A possibilistic fuzzy multi-objective approach was proposed and solved using an interactive fuzzy optimization solution methodology.\\ 
			
			\cite{Duan2019} & pulp and paper ind. & 5 & 2 products & -- & 20 & A combination of linguistic Z-numbers and alternative queuing method was used for supplier selection and multi-objective line programming was used for order allocation.\\ 
			
			\cite{Laosirihongthong2019} & cement manuf. & 15 & 3 materials & -- & 30 & FAHP was used for supplier evaluation and linear programming model for order allocation was solved using Microsoft solver.\\ 
			
			\cite{Ebrahim2019} & pharmaceutical & 6 & 3 products & -- & - & FAHP was used for supplier evaluation and multi-objective MILP model for order allocation was solved using CPLEX.\\ 
			

			\cite{Gupta2018} & numerical & 5 & -- & 4 plants, 6 warehouses, 8 retailers & 124 &A bi-level multi-objective model developed for SSOA was solved using fuzzy goal programming approach.\\ 
			
			\cite{Hu2018} & logistics & -- & & -- & 80 & A GA based approach was used to solve order allocation problem.\\ 
			

			\cite{Lo2018} & electronics & 6 & 6 products & 2 price levels & 144 & BWM, modified FTOPSIS was used for supplier evaluation and fuzzy multi-objective linear programming was used for order allocation.\\ 
			
			\cite{Mohammed2018} & meat supply chain & 7 & -- & 5 retailers & 34 & FAHP and FTOPSIS was used for supplier evaluation and the $\epsilon$-constraint method and LP-metrics method were employed to solve fuzzy multi-objective model for order allocation to get Pareto solution. Finally, TOPSIS was used to select final solution.\\ 
			
			\cite{Park2018} & bicycle & 13 & 6 components, 5 modules & -- & 143 & Multi-attribute utility theory was used for supplier region selection and weighted-sum method was used for solving SSOA.\\
			
			\cite{Mirzaee2018} & numerical & 3 & 3 products & 3 period & 207 & A preemptive fuzzy goal programming approach was used for solving MILP model of SSOA.\\
			
			\cite{Kim2018} & numerical & 10 & 5 items & 12 periods & 2436 & A branch-and-freeze algorithm, similar to branch-and-bound, was proposed to solve SSOA.\\ 
			

			\cite{Goren2018} & online retailer & 6 & 3 products & 3 period & 126 & A hybrid of fuzzy Decision Making Trial and Evaluation Laboratory and Taguchi Loss Functions were used for evaluating suppliers. Weighted comprehensive criterion method was used for solving bi-objective model for order allocation. \\ 
			
			\cite{Cheraghalipour2018} & plastic ind. & 5 & 3 items, 5 products & 5 periods,32 disruption scenarios, 2 types of quantity discount & 1630 & BWM was used for supplier evaluation and a Revised Multi-Choice Goal Programming method was used to solve order allocation.\\ 
			
			\multirow{2}{*}{\cite{Vahidi2018}} & auto. company & 4 & 6 parts & 14 disruption scenarios & 744 & Bi-objective allocation model was converted to single objective model using the weighted AEC method, and was solved using GAMS optimization solver and using a tailored differential evolution algorithm.\\ 
			& numerical & 15 & 10 parts & 10 disruption scenarios & 3450 & \\ 
			
			\cite{Shalke2018} & packaging ind. & 4 & 3 items & discount range 3, 6 periods & 204 & A multi-objective model for SSOA was solved using a multi-choice goal programming approach.\\ 
			
			\cite{Dupont2018} & numerical & 6 & -- & 8 customers & 2784 & An MILP model was used to show `elasticity of losses versus profits'.\\ 
			
			\cite{Aggarwal2018} & numerical & 5 & 3 products & 4 periods & 120 & A multi-objective model of the problem was solved using non-pre-emptive goal programming and weighted sum aggregate objective function technique.\\ 
			
			\cite{Bodaghi2018} & numerical & 10 & 10 parts & 10 periods, 5 orders, 4 contracted capacity intervals & 500 & FANP was used for supplier evaluation and new weighted fuzzy multi-objective model for SSOA and order scheduling was developed.\\
			
			\cite{Babbar2018} & beverages ind. & 6 & 2 products & 2 periods, 9 scenarios & 276 & A fuzzy quality function deployment was used for supplier evaluation and a stochastic multi-objective model for order allocation was solved using weighted-sums, $\epsilon$-constraint and distance approaches.\\ 
			
			\cite{Ghadimi2018} & electronics & 9 & 9 products & -- & 90 & A multi-agent system approach was developed for solving SSOA.\\ 
			

			\cite{Hamdan2017b} & numerical & 5 & -- & 3 periods, 4 discount ranges & 132 & AHP and FTOPSIS were used for supplier evaluation and a bi-objective model for order allocation was solved using MATLAB.\\ 

			\cite{Ghorabaee2017} & tissue paper manuf. & 5 & -- & & 10 & Interval type-2 fuzzy sets and the Evaluation based on Distance from Average Solution method are used for supplier evaluation and a multi-objective MILP for order allocation was solved using fuzzy programming approach.\\ 
			
			\cite{Bohner2017} & numerical & 10 & 15 products & 7 quantity discount intervals, 5 business volume discount intervals & 1,483,190 & An MILP model for SSOA was solved using exact optimization methods.\\ 
			
			\multirow{2}{*}{\cite{Hamdan2017}} & facilities management company & 3 & -- & 12 & 84 & AHP and FTOPSIS were used for supplier evaluation and a bi-objective and multi-objective models for order allocation was solved using weighted comprehensive criterion method. \\ 
			& numerical & 5 & -- & 6 periods & 66 & \\
			
			\cite{Rezaee2017} & numerical & 5 & 5 parts, 5 products & 5 refurbishing sites, 3 discount intervals & 240 & A multi-objective model for SSOA, based on the integrated simultaneous data envelopment analysis--Nash bargaining game, was solved using global criteria method.\\
			
			\cite{Kumar2017} & auto. company & 4 & -- & & 4 & FAHP was used for supplier evaluation and fuzzy multi-objective model for order allocation was solved using Lingo 13.\\ 
			
			\cite{Bakeshlou2017} & numerical & 3 & -- & & 3 & Fuzzy ANP (FANP) was used for supplier weights and FANP and fuzzy multi-objective linear programming model for order allocation was solved using weighted max-min method.\\ 
			
			
			
			
			
			
			
			\textbf{Our study} & \textbf{manuf.} & \textbf{70} & \textbf{5000} & double-allocation, penalty constraint, real-time application & \textbf{7,200,020} & Exact optimization and GA based approaches to solve SSOA problems.\\ \hline
	\end{longtable}
	\end{landscape}}
	
	We summarize the last five years' research literature on the SSOA problem, highlighting the supply chain size and number of variables in the optimization problem, in Table~\ref{tab_scale}. Due to missing parameters, a number of variables can not be calculated in a few references so they are left blank.
	
	It is clear that less attention has been given to real case studies and to large-scale numerical studies. Those that have considered real-life use cases did not integrate penalty discounts and did not consider iterative bidding processes which often characterize real-life use cases in industry. Bidding is an important aspect of SSOA, whereby suppliers are called into an conference/auction and submit bids to win a contract. While straightforward SSOA involves selecting and allocating suppliers at a single point in time, iterative bidding process brings uncertainty to the process, as suppliers cannot immediately be allocated based on price offering. Iterative bids, when done in real time, necessitates approaches that are fast and computationally efficient. The challenges brought by real-time bidding has not been studied well in SSOA problems. 
	
	We fill this gap in the literature by presenting a large-scale SSOA case study of a manufacturing company and studying a number of methods for solution approximation. This paper also presents a novel problem under the SSOA class of problems where order selection and allocation are considered on both the tiers of a 2-tier supply chain (resulting in cooperation) with multi-to-multi relationships in the items of two tiers, penalty constraints and real-time solution constraints. Furthermore, supplier preferences to work with other suppliers are considered through bidding in the problem formulation, presenting an advance to the current state of SSOA studies.

	\section{Problem description and formulation}
	\label{sec_problem}
	This section discusses problem details and formulation of optimization models.
	
	\subsection{Problem description}
	\label{subsec_description}
	The manufacturing company under consideration produces two types of engines: new and old engines, which require two types of parts, viz., blue-chips (parts which are required in currently manufactured engines) and LLV (low cost- low volume: parts which are required in legacy engines -- not manufactured currently) parts. The company needs estimated quantities of blue-chips and LLV parts to be used in different engines, which it outsources from a set of certified machining suppliers (hereon called Tier1). The suppliers are already selected according to some criteria (e.g., quality, service, social, environmental etc.) and it is assumed that there is no significant difference in the quality of items supplied by these suppliers. The demand for parts is known and constant, which is assigned to Tier1 suppliers under some constraints. After receiving orders for parts (blue-chips and LLV), Tier1 suppliers need different types of forged metal to manufacture the final finished parts. So, each Tier1 supplier outsources forgings to suppliers (hereon called Tier2). All parts and forgings have to be dual-sourced to avoid monopoly of items by the suppliers and to reduce the risk of disruptions in the supply chain (\cite{Tomlin2006}).
	
	The procurement of finished parts from Tier1 and forgings from Tier2 are carried out in sourcing auctions/conferences, where suppliers submit their bids as to the price with which they can offer to supply each item with associated transportation costs. The bidding process supports supplier preferences to work with other suppliers unlike the existing literature. After initial allocation from the bids, then there are multiple rounds of bidding, negotiations and allocations during the conference to obtain the best possible allocation. Because of the iterative nature of the bidding process, the company needs a dynamic model which could do assignments in real time, during each bidding round in the conference. During different rounds of bidding, only data changes but the problem structure remains the same, i.e., the model does not change. Thus, the iterative nature of the problem with time constraints requires the model to perform allocations in reasonable time to facilitate multiple rounds in the sourcing conference.
	
	Each supplier has limited capacity so there are some budget constraints on each of the suppliers. Additionally, each supplier can only supply a subset of the required items. Moreover, the company prefers some items to be procured by certain suppliers due to a variety of reasons such as the longevity of their relationship, and the service quality of the supplier. Since LLVs are low in number if a supplier does not get sufficient orders for blue-chips, it might incur extra cost to supply LLVs. Hence, if a Tier2 supplier does not get blue-chip orders greater than or equal to a certain amount from Tier1 suppliers then the Tier2 supplier increases the price of LLVs by a given factor, which is referred to as a penalty constraint.
	
	\begin{figure}[htb!]
		\centering
		\includegraphics[width=0.5\linewidth]{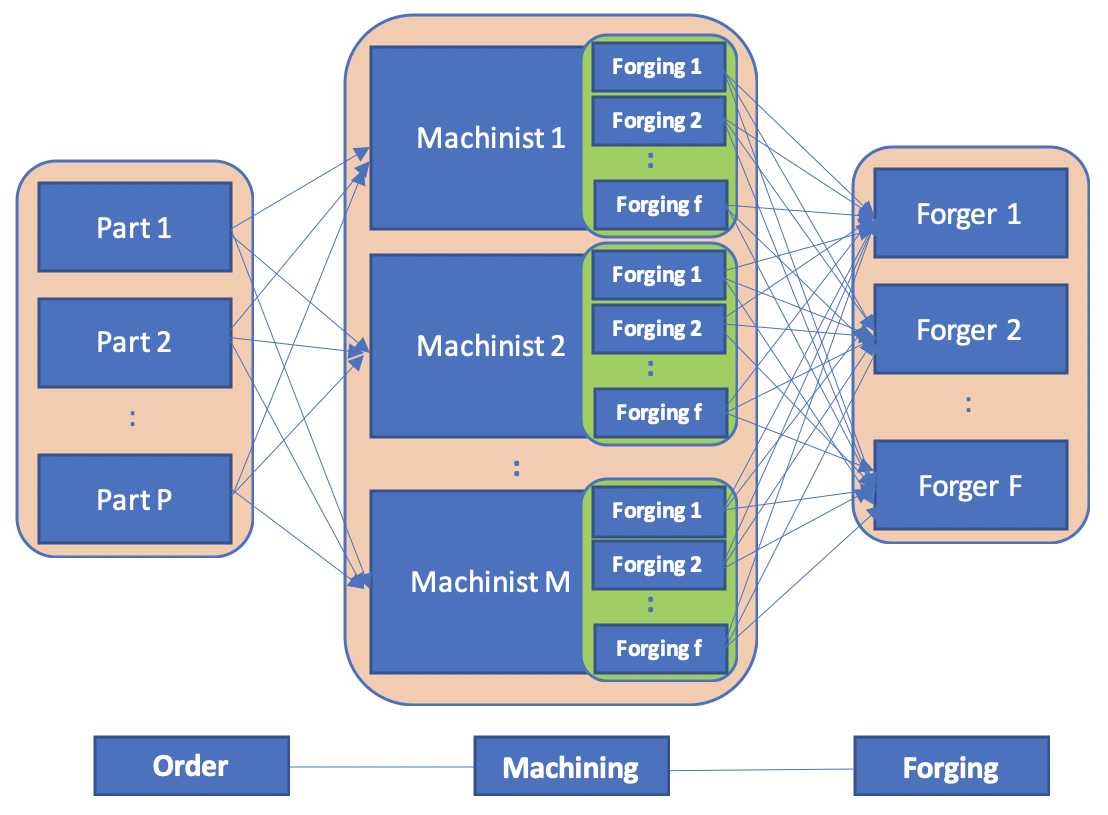}
		\caption{2-tier supply chain of a manufacturing company}
		\label{fig_SC}
	\end{figure}
	
	Thus, the company has a two-tier supply chain, as shown in the lower part of Fig.~\ref{fig_SC}, having machining operations and forging operations as two tiers. Forging involves manufacturing roughly shaped parts from melted alloys and machining refines those into final finished parts. The supply chain has two assignments: parts to Tier1 and forgings to Tier2, resulting in three alternative assignment problems: machinist-tier (considers assignment of parts to Tier1 only), forger-tier (considers assignment of forgings to Tier2 only) and an integrated problem (considers assignments of parts and forgings together). The integrated problem, involving the procurement from both the tiers, also facilitates cooperation between the tiers of the supply chain. Upper part of Fig.~\ref{fig_SC} presents the supply chain of the company where P parts are assigned to M Tier1 suppliers which then assign requirements of each of the forgings to F Tier2 suppliers. The overall procurement cost in the supply chain is the sum of costs of supplying forgings to Tier1 by Tier2 and costs of machining for manufacturing finished parts by Tier1.

	Presently, the manufacturer does not have any tool to solve any of the three order allocation problems (i.e., forger-tier problem, machinist-tier problem and integrated problem) due to the complexity of the problem. The allocations of orders is performed manually, by experienced procurement managers, and by organising separate sourcing conferences for assignment of parts and for assignment of forgings. Thus, the current practice followed by the manufacturer limits number of bidding rounds during the sourcing conferences, results in suboptimal allocations during each bidding round due to manual allocations, and does not allow cooperation between forging and machining tiers due to separate sourcing conferences. The presented model, as discussed below, solves all these issues.
	
	The objective of the problem is to design the supply chain of the company to reduce the overall procurement cost for parts across the supply chain by efficiently selecting and assigning orders to suppliers in both the tiers under the constraints given below.
	\begin{itemize}
		\item Each part and each forging must be dual-sourced. That means two suppliers will supply each item in a pre-defined proportions. This helps to avoid monopoly of supplier and in reducing the disruption risks.
		\item Each supplier (in both tiers) has upper and lower budget limits. That means although suppliers can submit bids for a large number of items but suppliers can not be assigned orders more than a pre-determined budget limits. Similarly, to maintain relationship with a supplier, it should be assigned some items equal to a pre-determined budget.
		\item Some Tier2 suppliers must supply certain forgings and some Tier1 suppliers must supply certain finished parts. We shall refer to this as a must-make constraint hereon. The runtime of solution approaches depend on the number of must-make constraints, which depends on the data, because an increase in must-make constraints reduces the decision space.
		\item Some Tier2 suppliers can not supply certain forgings and some Tier1 can not supply certain finished parts. That means either those items are not in the portfolio of the suppliers or manufacturer does not want to order those items to suppliers due to some reasons which further means that these suppliers do not bid for all items. Similar to must-make constraints, this constraints affects the runtime of solution approaches due to reduction in the decision space.
		\item The problem should be solved in real-time. This is because order allocations to suppliers are performed during sourcing conferences in multiple rounds of bidding. So, real-time solution is needed to facilitate the conferences.
		\item If a forging supplier does not get blue-chip orders greater than or equal to a certain amount then the forging supplier will increase the price of LLVs by some factor; or in other words, if a forger gets blue-chip orders greater than or equal to certain amount then the forger will reduce the price of LLVs by some factor. This results in supplier penalty to the buyer which is kind of opposite of supplier discounts to the buyer.
	\end{itemize}
	
	\subsection{Assumptions}
	\label{subsec_assumptions}
	In the development of model for the SSOA problem, the study assumes following points:
	\begin{itemize}
		\item The demand for parts is known and constant.
		\item Each supplier has limited capacity, imposed indirectly using budget limits.
		\item A set of certified Tier1 and Tier2 suppliers are already available, and satisfied certain a priori criteria to have become an approved supplier of the company, such as service quality, and compliance to social and environment laws.
		\item There is no significant difference in the quality of parts supplied by different Tier1 and forgings supplied by different Tier2. This is true to some extent because orders are assigned to certified suppliers.
		\item As such, consequent supplier selection and order allocation is done on the basis of unit price and unit transportation cost.
		\item Each supplier can provide a variety of items.
		\item There is a multi-to-multi relationship between forgings and parts, where one forging can be used to manufacture more than one part and similarly, one part can be manufactured from more than one forging.
	\end{itemize}
	
	\subsection{Problem formulations}
	\label{subsec_formulations}
	This subsection formulates two types of problems described earlier: single-tier problem which selects and allocates orders in a single-tier, i.e., either machinist-tier or forger-tier, and the integrated problem which selects and allocates orders across both the tiers simultaneously. Notational nomenclature is defined in Table~\ref{tab_notations}.
	
	{\small
		\begin{longtable}[H]{p{0.15\linewidth} p{0.85\linewidth}}
			\caption{Notation nomenclature}
			\label{tab_notations}\\
					\hline
					\textbf{Symbol} & \textbf{Meaning} \\
					\hline
					\endhead
					\multicolumn{2}{l}{\textbf{Indexes}}\\
					$i_B$/$i_L$ & index into parts (Blue chip)/ (LLV) \\
					$k_B$/$k_L$ & index into forgings (Blue chip)/ (LLV)\\
					$j$ & index into Tier1 \\
					$l$ & index into Tier2 \\
					$d$ & index into penalty factors \\
					\multicolumn{2}{l}{\textbf{Parameters}}\\
					$\gamma^d_l$ & penalty factor for forger $l$ at penalty $d$ where $\gamma^d_l$=1 for d=1 and $\gamma^d_l>1$ for d=2.\\
					$CBX_{i,j}$ & per unit machining cost for part $i$ by Tier1 supplier\\
					$CTX_{i,j}$ & per unit transportation cost for part $i$ by Tier1 supplier\\
					$CBY_{k,j,l}$ & per unit cost for supplying forging $k$ to Tier1 supplier j by Tier2 supplier l\\
					$CTY_{k,j,l}$ & per unit transportation cost for supplying forging $k$ to Tier1 supplier j by Tier2 supplier l\\
					$CX^{1}_{i,j}$/$CX^{2}_{i,j}$ & total cost of machining parts $i$ by Tier1 supplier for first/second proportion of dual-sourcing\\
					NpiB/NpiL & number of parts $i_B$/$i_L$ ordered\\
					$NX_{i}^1$/$NX_{i}^2$ & number of parts $i$ ordered in the first/second proportion of dual-sourcing\\
					s$_{i}$ & first proportion of the dual-sourcing for part $i$ so that (1-s$_{i}$) gives the second proportion\\
					s$_{k}$ & first proportion of the dual-sourcing for forging $k$ so that (1-s$_{k}$) gives the second proportion \\
					$S_{mj}^{min}$/$S_{mj}^{max}$ & lower/upper budget limit for Tier1 supplier\\
					$S_{fl}^{min}$/$S_{fl}^{max}$ & lower/upper limit on budget of Tier2 supplier l\\
					$y_{i,k}$ & number of forgings $k$ required to manufacture one unit of part $i$.\\
					$D_l$ & penalty threshold for Tier2 supplier l on blue-chips; if Tier2 supplier l gets blue-chip forging orders greater than or equal to $D_l$ then penalty $d=1$ and $\gamma^d_l = 1$ else $d=2$ and $\gamma^d_l > 1$\\
					$M$/$\epsilon$ & a very large/small number\\
					\multicolumn{2}{l}{\textbf{Functions}}\\
					$Z_{kj}$ & is a function of machinist assignments and returns number of forgings $k$ required by Tier1 supplier\\
					$CY^{1}_{k,j,l}$/$CY^{2}_{k,j,l}$ & is a function of machinist assignments and returns total cost of supplying forgings $k$ to Tier1 supplier by Tier2 supplier l for first/second proportion of dual-sourcing\\
					$CY^{1}_{k,j,l,d}$/$CY^{2}_{k_L,j,l,d}$ & is a function of machinist assignments and returns total cost of supplying forgings $k$ to Tier1 supplier by Tier2 supplier l at penalty d for first/second proportion of dual-sourcing\\
					\multicolumn{2}{l}{\textbf{Variables}}\\
					$X^{1}_{i_B,j}$/$X^{2}_{i_B,j}$ & 1 if part $i_B$ is supplied by Tier1 supplier j for first/second proportion of dual-sourcing otherwise 0\\
					$X^{1}_{i_L,j}$/$X^{2}_{i_L,j}$ & 1 if part $i_L$ is supplied by Tier1 supplier j for first/second proportion of dual-sourcing otherwise 0\\
					$Y^{1}_{k_B,j,l}$/$Y^{2}_{k_B,j,l}$ & 1 if forging $k_B$ supplied to Tier1 supplier j by Tier2 supplier l for first/second proportion of dual-sourcing otherwise 0\\
					$Y^{1}_{k_L,j,l,d}$/$Y^{2}_{k_L,j,l,d}$ & 1 if forging $k_L$ supplied to Tier1 supplier j by Tier2 supplier l at penalty d for first/second proportion of dual-sourcing otherwise 0\\
					$v_l$ & an indicator variable for penalty for Tier2 supplier l\\
					\hline
		\end{longtable}}

		Here, we calculate total cost of machining and forging for a specific part, proportion and supplier, which are common across different single-tier and integrated models. The machining cost can be pre-calculated as all the required information is available before hand but forging cost is dependent on Tier1 allocations, as given below. Let $CX_{i_B,j}^{1}$ is total cost of machining part $i_B$ by Tier1 supplier for first proportion of dual-sourcing ratio, which can be pre-calculated as (per unit machining cost for part $i_B$ + per unit transportation cost) $\times$ number of units of part $i_B$ supplied by Tier1 supplier for first proportion of dual-sourcing. This is because of the availability of all required parameters to calculate $CX_{i_B,j}^{1}$, which results in simplification of the model, as discussed in \ref{subsubsec_formulation}, and is given below.
		\begin{equation}
			CX_{i_B,j}^{1} = \left(CBX_{i_B,j} + CTX_{i_B,j}\right) \times NX_{i_B}^1,
		\end{equation}
		where $CBX_{i_B,j}$ is per unit machining cost, $CTX_{i_B,j}$ is per unit transportation cost, $NX_{i_B}^1 = s_{i_B} \times Npi_B$ and $Npi_B$ is the number of units of part $i_B$ ordered, and $s_{i_B}$ is the first proportion of dual-sourcing ratio for part $i_B$.
		Similarly, we can calculate costs $CX_{i_B,j}^{2}, CX_{i_L,j}^{1},\text{ and } CX_{i_L,j}^{2}$ for second proportion (by replacing superscript 1 with 2) and costs for part $i_L$ (by replacing $i_B$ with $i_L$). Let $CY_{k_B,j,l}^{1}$ is total cost of forging $k_B$ supplied by Tier2 supplier $l$ to Tier~1 supplier $j$ for first proportion and $CY_{k_L,j,l,d}^{1}$ is total cost of forging $k_L$ supplied by Tier2 supplier $l$ to Tier~1 supplier $j$ at penalty factor $\gamma^d_l$ for first proportion, are given below.
		\begin{equation}
			CY_{k_B,j,l}^{1} = \left(CBY_{k_B,j,l} + CTY_{k_B,j,l}\right) \times Z_{k_B,j}\times s_{k_B},
		\end{equation}
		\begin{equation}
			CY_{k_L,j,l,d}^{1} = \left(CBY_{k_L,j,l} \times \gamma^d_l + CTY_{k_L,j,l}\right) \times Z_{k_L,j} \times s_{k_L},
		\end{equation}
       where $CY_{k_B,j,l}/CBY_{k_L,j,l} \text{ and } CTY_{k_B,j,l}/CTY_{k_L,j,l}$ are per unit cost and per unit transportation cost for forging by Tier2 supplier $l$ for Tier1 supplier $j$, $Z_{k_B,j}/Z_{k_L,j}$ is the quantity of forgings ordered by $j$ from $l$ (calculated below) and $s_{k_B}/s_{k_L}$ is first proportion. 
		\begin{equation}
			\begin{split}
				Z_{k_B,j} = \sum_{i_B}\left(X_{i_B,j}^{1} \times y_{i_B,k_B} \times s_{i_B} \times NpiB +X_{i_B,j}^{2} \times y_{i_B,k_B} \times (1-s_{i_B}) \times NpiB \right) \\
				+ \sum_{i_L}\left(X_{i_L,j}^{1} \times y_{i_L,k_B} \times s_{i_L} \times NpiL +X_{i_L,j}^{2} \times y_{i_L,k_B} \times (1-s_{i_L}) \times NpiL \right),
			\end{split}
		\end{equation}
		where $y_{i,k}$ = yield, i.e., number of forgings $k$ required to manufacture one unit of part $i$, where each forging $k$ could be used to manufacture both the parts $i_B$ and $i_L$. Similarly, we can calculate $CY_{k_B,j,l}^{2}, CY_{k_L,j,l,d}^{2}$ and $Z_{k_L,j}$.
		
		Next, we define different optimization models for order assignments with dual-sourcing as follows.
		
		\paragraph{\textbf{Machinist-tier problem with dual-sourcing ($M_D$)}:} This problem involves selecting Tier1 suppliers to supply finished parts and allocating quantities of parts which are dual sourced. It considers only machinist-tier and optimizes the machining cost for finished parts, as given below.
		\begin{equation}
			\label{eq_M_D}
			\min _{X} \quad \sum_{i_B,j} \left( X_{i_B,j}^{1} \times CX_{i_B,j}^{1} + X_{i_B,j}^{2} \times CX_{i_B,j}^{2}\right) + \sum_{i_L,j} \left( X_{i_L,j}^{1} \times CX_{i_L,j}^{1} + X_{i_L,j}^{2} \times CX_{i_L,j}^{2}\right),
		\end{equation}
		subject to constraints,
		\begin{equation}
			\label{eq_M_D_dualB1}
			\sum_{j} X_{i_B,j}^{1} = 1, \quad \forall i_B	
		\end{equation}
		\begin{equation}
			\label{eq_M_D_dualB2}
			\sum_{j} X_{i_B,j}^{2} = 1, \quad \forall i_B	
		\end{equation}
		\begin{equation}
			\label{eq_M_D_dualB3}
			X_{i_B,j}^{1} + X_{i_B,j}^{2} <= 1, \qquad \forall i_B,j
		\end{equation}															
		
		\begin{equation}
			\label{eq_M_D_dualL1}
			\sum_{j} X_{i_L,j}^{1} = 1, \quad \forall i_L	
		\end{equation}
		\begin{equation}
			\label{eq_M_D_dualL2}
			\sum_{j} X_{i_L,j}^{2} = 1, \quad \forall i_L	
		\end{equation}
		\begin{equation}
			\label{eq_M_D_dualL3}
			X_{i_L,j}^{1} + X_{i_L,j}^{2} <= 1, \qquad \forall i_L,j
		\end{equation}														
		
		\begin{equation}
			\label{eq_M_D_budget}
			S_{mj}^{min} \le \sum_{i_B,i_L} \left( X_{i_B,j}^{1} \times CX_{i_B,j}^{1} + X_{i_B,j}^{2} \times CX_{i_B,j}^{2} + X_{i_L,j}^{1} \times CX_{i_L,j}^{1} + X_{i_L,j}^{2} \times CX_{i_L,j}^{2} \right)  \le S_{mj}^{max}, \forall j,
		\end{equation}
		\begin{equation}
			\label{eq_M_D_mustB}
			X_{i_B,j}^{1} + X_{i_B,j}^{2} = 1, \qquad \text{for set of values of } \lbrace \left(i_B,j\right) \rbrace,
		\end{equation}
		\begin{equation}
			\label{eq_M_D_mustL}
			X_{i_L,j}^{1} + X_{i_L,j}^{2} = 1, \qquad \text{for set of values of } \lbrace \left(i_L,j\right) \rbrace.
		\end{equation}
		Here, equation (\ref{eq_M_D}) is the objective function, which calculates the machining cost for finished parts as per the requirements. First part of objective function calculates machining cost for blue-chips and second part calculates the machining cost for LLVs. Equations (\ref{eq_M_D_dualB1}), (\ref{eq_M_D_dualB2}) and inequality constraints (\ref{eq_M_D_dualB3}) are used to force strict dual-sourcing of part $i_B$, where constraint (\ref{eq_M_D_dualB1}) ensures that only one supplier supplies the first proportion of dual sourcing for part $i_B$, constraint (\ref{eq_M_D_dualB2}) ensures that only one supplier supplies the second proportion of dual-sourcing for part $i_B$ and inequality constraint (\ref{eq_M_D_dualB3}) ensures that each supplier can supply part $i_B$ only for one proportion of dual-sourcing. Equations (\ref{eq_M_D_dualL1}), (\ref{eq_M_D_dualL2}) and inequality constraints (\ref{eq_M_D_dualL3}) are used to force strict dual-sourcing of part $i_L$, where constraint (\ref{eq_M_D_dualL1}) ensures that only one supplier supplies the first proportion of dual sourcing for part $i_L$, constraint (\ref{eq_M_D_dualL2}) ensures that only one supplier supplies the second proportion of dual-sourcing for part $i_L$ and inequality constraint (\ref{eq_M_D_dualL3}) ensures that each supplier can supply part $i_L$ only for one proportion of dual-sourcing. Inequality constraint (\ref{eq_M_D_budget}) forces budget constraints, where $S_{mj}^{min}$ and $S_{mj}^{max}$ are the lower and upper budget limits on Tier1 supplier j, respectively, and equality constraints (\ref{eq_M_D_mustB}) and (\ref{eq_M_D_mustL}) ensure must-make constraints on Tier1 suppliers for parts $i_B$ and $i_L$, respectively. To enforce the constraint, `some suppliers must not supply certain parts', we can have constraints similar to (\ref{eq_M_D_mustB}) and (\ref{eq_M_D_mustL}) with the right hand side being equal to zero. However, in order to keep the model simpler, we enforce these constraints by allocating large corresponding costs. 
		
		$M_D$ amounts to a mixed-integer linear programming problem.
		
		\paragraph{\textbf{Forger-tier problem with dual-sourcing ($F_D$)}:} This problem involves selecting Tier2 suppliers to supply forgings to Tier1, and allocating quantities of forgings with dual-sourcing. It considers only the forger-tier and optimizes the cost of supplying forgings by Tier1 to Tier2, as given below.
		\begin{equation}
			\label{eq_F_D}
			\min_{Y}  \sum_{k_B,j,l} \left(Y_{k_B,j,l}^{1} \times CY_{k_B,j,l}^{1} + Y_{k_B,j,l}^{2} \times CY_{k_B,j,l}^{2} \right) + \sum_{k_L,j,l,d} \left(Y_{k_L,j,l,d}^{1} \times CY_{k_L,j,l,d}^{1} + Y_{k_L,j,l,d}^{2} \times CY_{k_L,j,l,d}^{2} \right)
		\end{equation}
		\begin{equation}
			\label{eq_F_D_dualB1}
			\sum_{l} Y_{k_B,j,l}^{1} = 1, \quad \forall k_B,j
		\end{equation}
		\begin{equation}
			\label{eq_F_D_dualB2}
			\sum_{l} Y_{k_B,j,l}^{2} = 1, \quad \forall k_B,j	
		\end{equation}
		\begin{equation}
			\label{eq_F_D_dualB3}
			Y_{k_B,j,l}^{1} + Y_{k_B,j,l}^{2} <= 1, \quad \forall k_B,j,l
		\end{equation}		
		
		\begin{equation}
			\label{eq_F_D_dualL1}
			\sum_{l,d} Y_{k_L,j,l,d}^{1} = 1, \quad \forall k_L,j
		\end{equation}
		\begin{equation}
			\label{eq_F_D_dualL2}
			\sum_{l,d} Y_{k_L,j,l,d}^{2} = 1, \quad \forall k_L,j	
		\end{equation}
		\begin{equation}
			\label{eq_F_D_dualL3}
			\sum_{d} Y_{k_L,j,l,d}^{1} + Y_{k_L,j,l,d}^{2} <= 1, \quad \forall k_L,j,l
		\end{equation}	
		\begin{equation}
			\label{eq_F_D_budget}
			S_{fl}^{min} \le \sum_{k_B,j} \left(Y_{k_B,j,l}^{1} \times CY_{k_B,j,l}^{1} + Y_{k_B,j,l}^{2} \times CY_{k_B,j,l}^{2} \right) + \sum_{k_L,j,d} \left(Y_{k_L,j,l,d}^{1} \times CY_{k_L,j,l,d}^{1} + Y_{k_L,j,l,d}^{2} \times CY_{k_L,j,l,d}^{2} \right) \le S_{fl}^{max} \forall l
		\end{equation}
		\begin{equation}
			\label{eq_F_D_mustB}
			Y_{k_B,j,l}^{1} + Y_{k_B,j,l}^{2} = 1, \quad \text{for set of values of } \lbrace \left(k_B,j,l\right) \rbrace
		\end{equation}
		\begin{equation}
			\label{eq_F_D_mustL}
			\sum_{d} Y_{k_L,j,l,d}^{1} + Y_{k_L,j,l,d}^{2} = 1, \qquad \text{for set of values of } \lbrace \left(k_L,j,l\right) \rbrace
		\end{equation}
		\begin{equation}
			\label{eq_F_D_penalty1}
			-M v_l \le \sum_{k_B,j} \left(Y_{k_B,j,l}^{1} \times CY_{k_B,j,l}^{1} + Y_{k_B,j,l}^{2} \times CY_{k_B,j,l}^{2} \right) -D_l +\epsilon \le M(1-v_l), \quad \forall l
		\end{equation}
		\begin{equation}
			\label{eq_F_D_penalty2}
			\sum_{k_L,j} \left(Y_{k_L,j,l,d_1}^{1} \times CY_{k_L,j,l,d_1}^{1} + Y_{k_L,j,l,d_1}^{2} \times CY_{k_L,j,l,d_1}^{2} \right) \leq M(1-v_l), \quad \forall l
		\end{equation}
		\begin{equation}
			\label{eq_F_D_penalty3}
			\sum_{k_L,j} \left(Y_{k_L,j,l,d_2}^{1} \times CY_{k_L,j,l,d_2}^{1} + Y_{k_L,j,l,d_2}^{2} \times CY_{k_L,j,l,d_2}^{2} \right) \leq Mv_l \quad \forall l.
		\end{equation}
		
		Here, equation (\ref{eq_F_D}) is the objective function, which calculates the cost of supplying forgings to Tier1 as per the requirements. First part of objective function calculates cost of supplying blue-chip forgings and second part calculates the cost of supplying LLV forgings. Equations (\ref{eq_F_D_dualB1}), (\ref{eq_F_D_dualB2}) and inequality constraint (\ref{eq_F_D_dualB3}) are used to force strict dual-sourcing of forgings $k_B$, 
		where constraint (\ref{eq_F_D_dualB1}) ensures that only one supplier supplies the first proportion of dual sourcing for part $k_B$, constraint (\ref{eq_F_D_dualB2}) ensures that only one supplier supplies the second proportion of dual-sourcing for part $k_B$ and inequality constraint (\ref{eq_F_D_dualB3}) ensures that each supplier can supply part $k_B$ only for one proportion of dual-sourcing.
		Equations (\ref{eq_F_D_dualL1}), (\ref{eq_F_D_dualL2}) and inequality constraint (\ref{eq_F_D_dualL3}) are used to force strict dual-sourcing of forgings $k_L$,
		where constraint (\ref{eq_F_D_dualL1}) ensures that only one supplier supplies the first proportion of dual sourcing for part $k_L$, constraint (\ref{eq_F_D_dualL2}) ensures that only one supplier supplies the second proportion of dual-sourcing for part $k_L$ and inequality constraint (\ref{eq_F_D_dualL3}) ensures that each supplier can supply part $k_L$ only for one proportion of dual-sourcing.
		Inequality constraint (\ref{eq_F_D_budget}) forces budget constraints, where $S_{fl}^{min}$ and $S_{fl}^{max}$ are lower and upper budget limits on Tier2 supplier l, respectively, and equality constraints (\ref{eq_F_D_mustB}) and (\ref{eq_F_D_mustL}) ensure must-make constraints on Tier2 for forgings $k_B$ and $k_L$. Inequality constraints (\ref{eq_F_D_penalty1}) to (\ref{eq_F_D_penalty3}) are used to ensure penalty constraints with the help of indicator variable $v_l$ which is 1 if blue-chip forging allocation to Tier2 supplier l is below $D_l$ else 0.
		
		$F_D$ is also a mixed-integer linear programming problem.
		
		\paragraph{\textbf{Integrated problem with dual-sourcing ($I_D$)}:} This problem considers both machinist and forger tiers and optimizes the overall cost of the supply chain for selecting Tier1 suppliers and allocating orders for parts; and selecting Tier2 suppliers and allocating orders for forgings. Hence, the total procurement cost of supply chain is the sum of machining cost for parts by Tier1 suppliers and the cost for supplying forgings by Tier2 suppliers to Tier1, which is given as follows.
		\begin{equation}
			\begin{split}
				\min_{X, Y} \sum_{i_B,j} \left( X_{i_B,j}^{1} \times CX_{i_B,j}^{1} + X_{i_B,j}^{2} \times CX_{i_B,j}^{2}\right) + \sum_{i_L,j} \left( X_{i_L,j}^{1} \times CX_{i_L,j}^{1} + X_{i_L,j}^{2} \times CX_{i_L,j}^{2}\right) \\
				+ \sum_{k_B,j,l} \left(Y_{k_B,j,l}^{1} \times CY_{k_B,j,l}^{1} + Y_{k_B,j,l}^{2} \times CY_{k_B,j,l}^{2} \right) + \sum_{k_L,j,l,d} \left(Y_{k_L,j,l,d}^{1} \times CY_{k_L,j,l,d}^{1} + Y_{k_L,j,l,d}^{2} \times CY_{k_L,j,l,d}^{2} \right)
			\end{split}
		\end{equation}
		The constraints on the integrated problem are combined from constraints on machinist-tier and forger-tier problem, i.e., equations (\ref{eq_M_D_dualB1}), (\ref{eq_M_D_dualB2}), (\ref{eq_M_D_dualL1}), (\ref{eq_M_D_dualL2}), (\ref{eq_F_D_dualB1}), (\ref{eq_F_D_dualB2}), (\ref{eq_F_D_dualL1}), (\ref{eq_F_D_dualL2}) and inequalities (\ref{eq_M_D_dualB3}), (\ref{eq_M_D_dualL3}), (\ref{eq_F_D_dualB3}) and (\ref{eq_F_D_dualL3}) are used to force strict dual-sourcing. Inequalities (\ref{eq_M_D_budget}) and (\ref{eq_F_D_budget}) force upper and lower budget constraints on Tier1 and Tier2, respectively, and equalities (\ref{eq_M_D_mustB}), (\ref{eq_M_D_mustL}), (\ref{eq_F_D_mustB}) and (\ref{eq_F_D_mustL}) are used to ensure must-make constraints on Tier1 and Tier2, respectively. Inequalities (\ref{eq_F_D_penalty1}) to (\ref{eq_F_D_penalty3}) are used to ensure penalty constraints. $I_D$ is a mixed-integer quadratic linear programming problem with quadratic constraints because forger-tier is dependent on machinist-tier as number of forgings requirements by each Tier1 supplier is dependent on the assignments of parts to the supplier.

		\section{Solution approaches}
		\label{sec_solution}
		This section discusses two different approaches for solving the problems discussed in the previous section: Mathematical Programming (MP) and Genetic Algorithms (GA). Because MP based methods can guarantee optimal solutions to an optimization problem so they are our first choice (\cite{Chauhan2022exploitation,chauhan2022trolley}). However, these methods might not be able to solve complex problems. On the other hand, meta-heuristic methods, like GAs are suitable for solving some of the complex problems, although, they can not guarantee optimality. Moreover, GA is the most widely used meta-heuristic method to solve SSOA problem (\cite{Aouadni2019}). Due to these reasons, MP and GA based methods are selected to solve the SSOA problem described in this paper.
		
		We will present how these particular solution approaches can be most effectively applied to the problem at hand, so as to take into account the industrial requirements of minimal run time during the auctioning/sourcing conference process, whilst achieving an optimal solution as possible within the highly constrained, large-scale solution space.
		
		\subsection{Mathematical programming}
		\label{subsec_mp}
		Single-tier problems (machinist-tier and forger-tier) described earlier are Mixed-Integer Linear Programming (MILP) problems that can be solved using MP to optimality. However, the integrated problem is a Mixed-Integer Quadratic Programming (MIQP) problem where objective as well as constraints are quadratic. Due to the quadratic nature and complexity of the supply chain, it is computationally infeasible to solve the integrated problem directly. Hence, if we were to use MILP in this problem, a 2-phase approach would need to be developed which divides the integrated problem into two phases. In phase one, the machinist-tier problem is solved and in phase two the forger-tier problem is solved using solution of phase one. Since the individual problems are linear, the problem can be reduced to a sequential linear problem. This approach helps to solve an approximate feasible solution to this complex problem but does not guarantee the optimal solution due to its sequential nature.
		
\begin{figure}[htb!]
			\centering
			\includegraphics[width=0.4\linewidth]{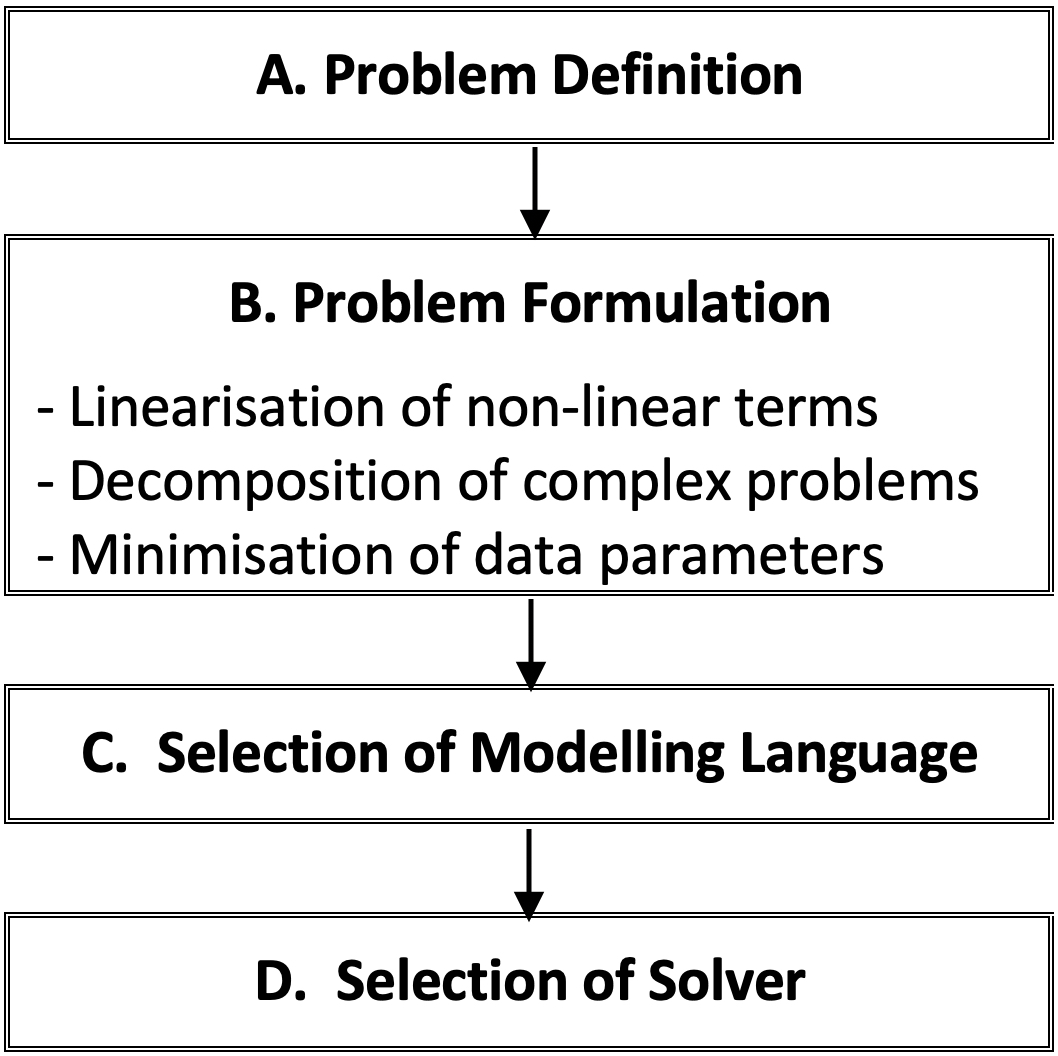}
			\caption{MP approach to solve problems}
			\label{fig_methodology}
		\end{figure}

		Fig.~\ref{fig_methodology} represents our approach to solve the assignment problems using MP, which has four steps, as discussed below.
		
		\subsubsection{Problem definition}
		\label{subsubsec_def}
		Problem definition is the first step in the problem solving where we understand the problem requirements, describe the problem environment, assumptions and expected usage of problem solution. This is an important step and the more time is spent on problem definition, the easier are the remaining steps. Section~\ref{subsec_description} and \ref{subsec_assumptions}, have defined the problem undertaken in the current paper.
		
		\subsubsection{Problem formulation}
		\label{subsubsec_formulation}
		Problem formulation involves identifying decision variables, finding relationships among decision variables and formulating objective function of problem. Decision variables are the entities which we want to find and form the solution of the problem. Relationships among variables translate to constraints in the optimization problem. This is a crucial step that determines computational complexity (\cite{chauhan2019problem}). For instance, discrete and continuous optimization problems need different approaches to solve them. Linear problems are easier to solve than non-linear optimization problems. Here, we discuss how this complex industrial problem could be simplified.
		
		\paragraph{\textbf{linearization:}} linearization is a technique that converts non-linear terms in an optimization problem to linear terms, hence simplifying the problem. However, linearization increases the number of decision variables and the number of constraints in the problem. For example, suppose we want to linearize $x \times y$, where $x$ and $y$ are binary decision variables. This could be done by introducing a new binary variable $z=x \times y$ and the following constraints.
		\begin{equation}
			\label{eq_linearise1}
			z \le x, \qquad z \le y,
		\end{equation}
		\begin{equation}
			\label{eq_linearise2}
			z \ge x + y -1.
		\end{equation}
		Here, inequality~(\ref{eq_linearise1}) ensures that $z$ is zero whenever any of $x$ or $y$ is zero, whereas (\ref{eq_linearise2}) ensures that $z$ is one if both $x$ and $y$ are one. For a general case of n binary variables, suppose we want to linearizethe product $ \prod_{i=1}^{n} x_i$. Then, we may introduce a binary variable $z$ such that,
		\begin{equation}
			\label{eq_linearise3}
			z \le x_i, \quad \text{for } i=1,2,3,...,n,
		\end{equation}
		\begin{equation}
			\label{eq_linearise4}
			z \ge \sum_{i=1}^{n} x_i - (n-1).
		\end{equation}
		In our case, the penalty constraint makes the order assignment problem a non-linear problem. To avoid this non-linear term, we introduced an extra dimension to the variables and cost parameters corresponding to the LLV forgings, i.e., to $Y_{k_L,j,l,d}$ and $CY_{k_L,j,l,d}$, respectively. Without linearization, taking $p_l \in \{\gamma^1_l, \gamma^2_l\}$ as a penalty variable, the original non-linear term would have been $Y_{k_L,j,l} \times CY_{k_L,j,l}$, where $CY_{k_L,j,l}$ is given below as:
		\begin{equation}
			\begin{split}
				CY_{k_L,j,l} = \left(CBY_{k_L,j,l} \times p_l + CTY_{k_L,j,l}\right) \times Z_{k_L,j}.
			\end{split}
		\end{equation}
		Note that the integrated problem can also be converted into a linear problem using similar techniques however the linearization results into billions of variables, which makes the problem computationally infeasible, which is why we follow a two phase approach. (Please refer to Appendix~\ref{app_lin} for linearization of integrated problem.)
		
		\paragraph{\textbf{Decomposition:}} Decomposition divides a complex problem into smaller problems, which are solved separately and solutions are combined to obtain the solution of original problem. The resulting problems are simpler than the original, and can be solved easily, however, the approach removes the guarantee of obtaining optimal solutions of the problem. In our case, the integrated SSOA problem is computationally infeasible so we could solve it by dividing into two phases. Here, the first phase solves the machinist-tier and second phase solves the forger-tier using solution of machinist-tier.
		Please refer to \cite{Boyd2007,Kinable2014} for more on decomposition approaches for optimization methods.
		
		\paragraph{\textbf{Simplification of data parameters:}} Wherever possible, all the data parameters (matrices) should be pre-computed and minimum number of data parameters should be used. This helps in reducing computational effort and thus run time. For example, in our case, cost matrices (like $CX_{i_B,j}^{1}$) are pre-computed to minimize the number of data parameters into the objective function. Similarly, all the constraints are enforced on the fly and no data parameters are used to enforce them.
		
		\subsubsection{Selection of modelling language}
		\label{subsubsec_modellingL}
		A modelling language makes it possible to solve a problem by plugging different solvers without changing the code itself. For instance, a problem written in Python MIP library can be solved with CBC, Gurobi or other supported solvers, by just specifying the solver. Each modelling language has its own features, and these need to be taken into account when choosing one that is suitable to the problem at hand. For instance, the Python interface of Gurobi, as a modelling language, allows for Python to perform fast matrix operations. Python MIP library and Pyomo support creation and reusability of expressions but neither supports faster matrix operations.
		Generally, modelling languages takes more than half of the total computation time (\cite{Lee2020}) and for large-scale problems, it could take huge amount of time to just generate a model. We observed that pyomo and python interface of Gurobi gets stuck with the model generation for integrated problem so Python MIP library is used which is also available freely.
		
		\subsubsection{Selection of solver}
		\label{subsubsec_solver}
		Solver libraries provide methods to actually solve the optimization problem. Gurobi, CPLEX and CBC are examples of popular solver libraries. Most of the solver libraries are implemented in C language and also use parallel computing, to provide faster computations, e.g., CBC and Gurobi. The performance of problem solving is also impacted by the selection of solver libraries. Moreover, the choice of solvers depend on their availability as open-source or commercial software and the problem nature etc. For example, CBC is freely available but can solve only linear and quadratic problems, on the other hand, Gurobi is a commercial software (with free academic access) and can solve a wide variety of problems. Thus, there is no single best solver and appropriateness of a solver depends upon the problem in hand (\cite{Anand2017}). For recent comparative studies on optimization solvers readers can refer to \cite{Anand2017,Lee2020,Wikarek2018}.

		\subsection{Genetic algorithms}
		\label{subsec_ga}
		Genetic Algorithms (GAs) are type of meta-heuristic random search method, based on Darwin's theory of natural evolution. In GAs, fittest individuals are selected from a population of solutions for reproduction of offspring solutions for the next generation (i.e., iteration). GAs start with a random population of solutions and move from one population to another using selection, crossover and mutation of solutions. The GA terminology and procedure are described below.
		
\begin{figure}[htb]
			\centering
			\includegraphics[width=0.7\linewidth]{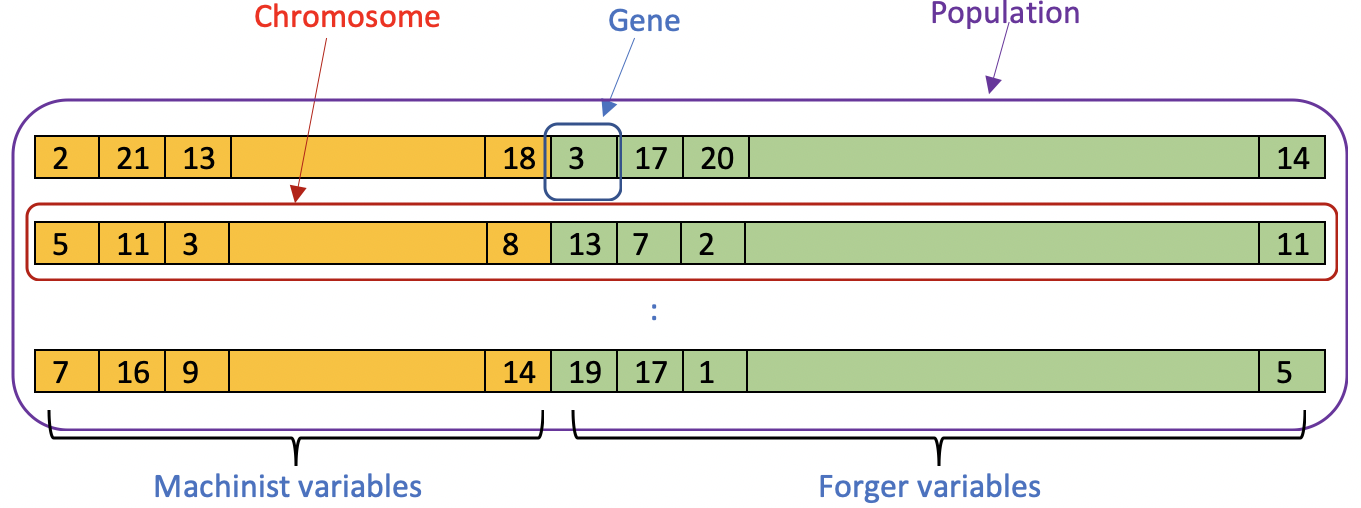}
			\caption{An example of gene, chromosome and population in the SSOA problem.}
			\label{fig_gene} 
		\end{figure}

		\textbf{Gene}: Refers to a single parameter/variable of an individual solution. In our problem, there are two variables for each part and each forging, which refer to suppliers for supplying two proportions of the part/forging. For example, a gene is highlighted in Fig.~\ref{fig_gene} where three means supplier three to supply that forging.
		
		\textbf{Chromosome}: Refers to an individual solution of the problem and is formed by joining the genes together. Allocation of parts to Tier1 and forgings to Tier2 form a solution and is a chromosome as shown in the Fig.~\ref{fig_gene}. Size of each chromosome is equal to number of variables which is equal to twice of sum of number of parts and number of forgings, where two accounts for the dual-sourcing. Since we have two types of allocations so we have used two layer encoding, represented by two colours in the chromosome where yellow colour refers to allocation of parts to Tier1 and green colour refers to allocation of forgings to Tier2.
		
		\textbf{Population}: Refers to a set of individual solutions or set of chromosomes, also called a generation. GA starts with a set of random allocations, called initial population/generation, and evolves from generation to generation using different GA operators, as discussed below.
		
		\textbf{Fitness function}: Refers to the function used to evaluate goodness of an individual solution, i.e., goodness of allocations to suppliers. Since goodness of allocation is dependent on the procurement cost across the supply chain so objective function of MIP models developed in previous subsection are used as fitness function.
		
		\textbf{Selection}: A genetic operator that is used to select the fittest individuals (parents) and let them pass their best genes to the next population and help in evolution of solution. There are several methods for selection such as tournament selection, roulette wheel selection and rank selection etc.
		
		\textbf{Crossover}: A genetic operator, also called recombination, that is used to combine two parents/solutions /allocations to form two new offspring/allocations. So, this is a way to stochastically generate new allocations for suppliers from the existing allocations of suppliers. There are several operators for crossover such as uniform crossover, one-point and two-point crossover etc.
		
		\textbf{Mutation}: A genetic operator that is used to change some of the genes with a low probability in certain new offspring, and is controlled through a hyperparameter called mutation probability. Mutation helps solution diversification to avoid premature convergence of GA. There are several methods for mutation such as uniform mutation, flip bit and Gaussian etc.
		
		In applying GA to our problem space, we need to keep in mind the large number of decision variables. Here, integer encoding is used as opposed to binary, so as to reduce the number of variables. Further, we use two layer encoding, where one layer is for Tier1 supplier allocations and the other is for the Tier2 supplier allocations, as described in the Fig.~\ref{fig_gene}. Each gene of the two layers refers to a supplier and number of genes in each layer is twice of number of items in that layer to account for dual-sourcing where one supplier supplies first proportion and second supplies second proportion of order for the item.
	
		The GA algorithm used to solve the problems is given below.
		
		\textbf{Algorithm 1.} Genetic Algorithm to solve Supplier Selection and Order Allocation Problem
		
		\textbf{Inputs}: Number of generations G, population size N, crossover probability $p_c$, mutation probability $p_m$, percentage of individuals to replace R.
		
		\textbf{Step 1.} Randomly generate a population of size N, i.e., a set of N allocations for parts to Tier1 and forgings to Tier2. The value of N is determined by trial, as discussed below.
		
		\textbf{Step 2.} For each individual solution, i.e., allocations to suppliers (for parts as well as for forgings), verify each of the problem constraints discussed in the model development, i.e., check if solution is feasible or not. If any of the solution, i.e., allocation of suppliers violates any of the problem constraint then randomly modify that part of the supplier allocation to produce a feasible solution.
		
		\textbf{Step 3.} Calculate fitness value for all individuals, i.e., calculate the procurement cost for all allocations of suppliers, using objective function of MIP models developed in previous subsection. Since the allocation problems are constrained optimization problems and we need to verify the constraints for checking the feasibility of the allocations to suppliers. So, we need to call Step 2 every time we calculate fitness function, i.e., before calling Step 3.
		
		\textbf{Step 4.} Apply selection operator to select individuals to generate new offspring, i.e., from population of allocations (allocations of parts to Tier1 and allocations of forgings to Tier2), select two allocations/solutions to pass the best allocation genes to the next generation. This selection is dependent on the fitness value, i.e., procurement cost of the allocations and there are several methods to achieve this. We employ tournament selection operator which selects the winners of the tournaments to generate new allocations.
		
		\textbf{Step 5.} Apply crossover on a pair of parents to generate a pair of children, i.e., pair of solutions/allocations obtained from previous step are stochastically combined to generate two new allocations. We employed uniform crossover with a probability of $p_c$ due to the large solution space of the problem.
		
		\textbf{Step 6.} Apply mutation with probability $p_m$ on each offspring, i.e., for each new allocation generated by crossover step, some of the allocation genes are randomly changed to increase the diversity in the allocations to avoid getting stuck in local minima. The value of $p_m$ is, generally, very low, i.e., only few allocation genes are changed otherwise best genes inherited from previous generation can be lost.
		
		\textbf{Step 7.} Repeat Steps 4 to 6 to generate R new individuals/allocations and update old generation with newly generated individuals to get new generation. We select N best allocations from the old and newly generated allocations which follows elitism as best allocations from previous set of allocations are carry forwarded to the new set of allocations.
		
		\textbf{Step 8.} Repeat Steps 2 to 9 for G generations. Each such repetition starts with population/generation, i.e., a set of allocations which are first verified for feasibility and then fitness scores are calculated for solutions. Those scores are then used in the selection process to select parents to generate new offspring which then undergo mutations to avoid local minima, and finally new set of allocations are selected using elitism from old and new allocations. The algorithm moves from generation to generation to generate the final allocations.
		
		A variety of operators can be used for selection, crossover and mutation with this algorithm to solve all the assignment problems. For choosing these operators and other hyperparameters of GA, we employ trial and error method, similar to \cite{Meena2013}, as detailed below.
		
		Given the inherent randomness to the problem, an experimental set up is conducted first, where all GA parameters are trialled over a large range: population size N (50 to 300), uniform cross-over probability $p_c$ (0.6 to 1.0), mutation-probability $p_m$ (0.0001 to 0.1), number of generations G (100 to 1000) and replacement of old generations R (90 to 100\%). Next, each parameter in the selected range is varied, keeping all other variables fixed, in the order: crossover probability $p_c$, mutation probability $p_m$, population size N, number of generations G and replacement percentage R of old generation. Then, previous step is repeated until results become stable, i.e., there is no more improvement by changing these parameters.
		
		After the set up experiments, following operators are selected for GA: uniform crossover, uniform mutation and tournament selection (with size 5) as crossover, mutation and selection operators, respectively. Moreover, GA uses 100, 0.001, 1.0, 90 and 50,000 as population size N, mutation probability ($p_m$), crossover probability ($p_c$), number of solutions to replace R and number of generations G, respectively.

		\section{Experimental results: case study}
		\label{sec_experiments}
		This section discusses the data, and experimental setup for solving the problem and the results.
		\subsection{Data and experimental setup}
		\label{subsec_data}
		The case study under consideration belongs to a supply chain of a manufacturing company and the models developed were deployed to select suppliers and allocate orders to them during a sourcing conference organized by the company. The problem size is represented in Table~\ref{tab_problem}.
		\begin{table}[htb!]
			\centering
			\caption{Problem size}
			\label{tab_problem}
			\begin{tabular}{|l|r|}
				\hline
				\#Tier2 & 20 \\ \hline
				\#Tier1 & 50 \\ \hline
				\#parts (Blue-chip) & 1500 \\ \hline
				\#parts (LLV) & 500 \\ \hline
				\#forgings (Blue-chip) & 2500 \\ \hline
				\#forgings (LLV) & 500\\ \hline
			\end{tabular}
		\end{table}
		There are 2000 types of parts of which 1500 belongs to blue-chip category and 500 to LLV. Each part is ordered in certain quantities. Further, each part is sourced according to the sourcing strategy from 50 different Tier1 suppliers. Each of the supplier in Tier1 then orders 3000 forgings (2500 blue-chip and 500 LLV) from 20 different suppliers in Tier2. The number of forgings required by each of the Tier1 suppliers depends on the part orders received by the supplier.
		
		Due to the confidentiality of the data, the data is not shared by the manufacturer and a representative case is generated by the OEM, as discussed here. Order for each parts (LLVs as well as blue-chips) is generated randomly between 100 and 500. Yield, i.e., number of forgings required to manufacture a part is randomly decided between one and three. For parts, per unit cost of machining is generated between 5000 and 10000 and per unit transportation cost is generated between 2 and 100. For forgings, per unit cost is generated between 1 and 10 and per unit transportation cost is generated between 1 and 5. For simplicity, we have put very loose upper and lower budget constraints as 0 and 1e+12. We introduced 5 must make constraints for Tier1 and Tier2 for both types of items, i.e., LLVs and Blue-chips. The penalty factor for non-preferred allocations is taken as 5 and penalty threshold is taken as 1000.
		
		The representation of a problem depends upon the approach used to solve the optimization problem. Table~\ref{tab_vars}
		\begin{table}[htb!]
			\centering
			\caption{\#Variables in MP and GA approaches (S: single-sourcing and D: dual-sourcing)}
			\label{tab_vars}
			\begin{tabular}{|c|r|r|}
				\hline
				\textbf{Problem} & \multicolumn{1}{c|}{\textbf{MP}} & \multicolumn{1}{c|}{\textbf{GA}} \\ \hline
				machinist-tier  (S) & 100,000 & 2,000 \\ \hline
				machinist-tier  (D) & 200,000 & 4,000 \\ \hline
				forger-tier  (S) & 3,500,020 & 150,000 \\ \hline
				forger-tier  (D) & 7,000,020 & 300,000 \\ \hline
				Integrated (S) & 3,600,020 & 152,000 \\ \hline
				Integrated (D) & 7,200,020 & 304,000 \\ \hline
			\end{tabular}
		\end{table}
		represents the number of variables in mathematical programming (MP) and in Genetic Algorithm (GA) approaches. The number of variables in the machinist-tier problem $M_D$ using MP can be calculated as $\left[ X^1_{i_Bj} \right] + \left[ X^2_{i_Bj} \right] + \left[ X^1_{i_Lj} \right] + \left[ X^2_{i_Lj} \right]$ = $\#parts (\text{blue-chips}) \times \#Tier1 + \#parts (\text{blue-chips}) \times \#Tier1+ \#parts (LLV) \times \#Tier1+ \#parts (LLV) \times \#Tier1$ = 1500$\times$50 + 1500$\times$50 + 500$\times$50 + 500$\times$50 = 200,000. 
		
		For the GA approach, the number of variables = 2 $\times$ \#parts (blue-chip) + 2 $\times$ \#parts (LLV) = 2 $\times$ 1500+2 $\times$ 500 = 4,000. Similarly, we can calculate number of variables in forger-tier and integrated problems. The large number of variables in the proposed problem formulation is mainly due to two reasons: first, unlike the existing literature, the presented case study needs allocations at two interdependent tiers of the supply chain and secondly, the presented supply chain is in itself complex due to relatively larger size (please refer to Table~\ref{tab_scale} for comparative study of size against the literature).
		
		All experiments are executed on a Linux Server (RAM 64 GB, 56 CPU) and averaged over five runs. For GA, PGAPack library\footnote{https://github.com/schlatterbeck/pgapack} was used which is a C based library. GA uses uniform crossover, uniform mutation and tournament selection (with size 5) as crossover, mutation and selection operators, respectively. Moreover, GA uses 100, 0.001, 1.0, 90 and 50,000 as population size, mutation probability ($p_m$), crossover probability ($p_c$), number of solutions to replace R and number of generations G, respectively. For MP, Python MIP (Mixed-Integer Linear Programming) library is used for modelling and Gurobi~9.0 is used as a solver library, which is written in C language.
		
		\subsection{Comparative study}
		\label{subsec_comparison}
		 Here, we discuss the convergence of the GA and comparative analysis of MP versus GA, and also comparison with the existing approach employed by the manufacturing company and the meta-heuristic algorithms.
		
\begin{figure}[htb!]
			\centering
			\includegraphics[width=0.5\linewidth]{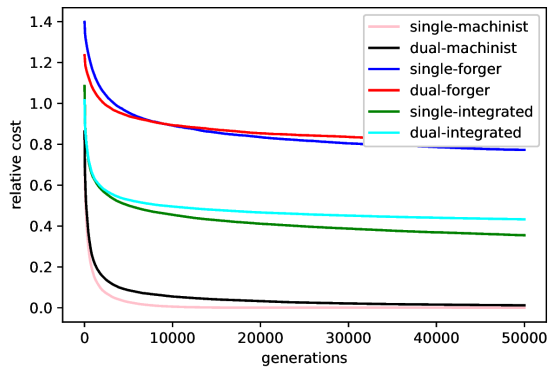}
			\caption{Convergence of GA algorithm for the three problems.}
			\label{fig_ga_result}
		\end{figure}

		Fig.~\ref{fig_ga_result} represents the convergence of GA for all the three problems where we plot relative procurement cost against number of generations. The relative cost is obtained by first subtracting and then dividing the procurement cost by the procurement cost obtained from the MP approach. From the figure, it is clear that single-sourcing problems (please refer to Appendix~\ref{app_single} for problem formulations and Appendix~\ref{app_singlesourcing_results} for results with the single-sourcing problems) converge to better solutions than dual-sourcing because of smaller number of variables. The machinist-tier problems converge to better solutions than others and single-sourcing problem manages to converge to the optimal solution.
		Moreover, integrated problems, which have more variables than single tier problems, converge to better solution than forger-tier problems. This is because integrated problem assigns orders to Tier1 and Tier2 simultaneously and can get better assignments, unlike forger-tier problem which has a fixed order pattern from the machinist tier. Another observation about the convergence of GA algorithms is that despite quickly getting closer to the solution, it takes large number of generations, and hence long time, to get a good solution.
		
		Table~\ref{tab_mp_vs_ga_dual} compares MP and GA approaches in terms of time (in minutes) to solve the problem and ratio of procurement cost and best procurement cost obtained. For experimental simplicity, we consider same dual-sourcing equal to 70:30 for all items.
		\begin{table}[htb!]
			\centering
			\caption{MP versus GA with dual-sourcing of 70:30}
			\label{tab_mp_vs_ga_dual}
			\begin{tabular}{|c|r|r|r|r|}
				\hline
				\multirow{2}{*}{\textbf{Problem}} & \multicolumn{2}{c|}{\textbf{MP}} & \multicolumn{2}{c|}{\textbf{GA}} \\ \cline{2-5} 
				& \multicolumn{1}{c|}{Time (minutes)} & \multicolumn{1}{c|}{Cost/best} & \multicolumn{1}{c|}{Time (minutes)} & \multicolumn{1}{c|}{Cost/best} \\ \hline
				\textit{machinist-tier} & 0.149 & 1.00000 & 94.977 & 1.01189 \\ \hline
				\textit{forger-tier} & 40.690 & 1.00000 & 7836.789 & 1.80964 \\ \hline
				\textit{Integrated} & 40.986 & 1.00000 & 11001.438 & 1.43275 \\ \hline
			\end{tabular}
		\end{table}
		
		From the table, it is clear that MP is better than GA both in terms of cost and time to solve the problem, including the integrated problem which is solved using a 2-phase approach by MP.
		Since GA approach to solve SSOA takes a very long time so it is not practical as it cannot solve the forger-tier and integrated problems in real-time which was a key requirement, i.e., it can not be used iteratively during multiple rounds of bidding process in the auction for order allocation. On the other hand, MP approach to solve SSOA is relatively much faster as it can solve machinist-tier allocation in less than a minute and can solve the integrated allocations in less than an hour which is practical to be used in real-time during the auction.
		
		It is interesting that MP approach solves all the three allocations in lesser time and produces better costs than GA for a complex problem because, generally, GAs are proposed as a more suitable to complex problems (\cite{Vishnu2021}). While this may indeed be the case in dealing with non-linear problems, however, GA is not the best option for solving large-scale SSOA problems as they are memory and computation intensive.
		\begin{figure*}[htb!]
			\centering
			\begin{subfigure}[t]{0.5\textwidth}			
				\centering
				\includegraphics[width=\textwidth]{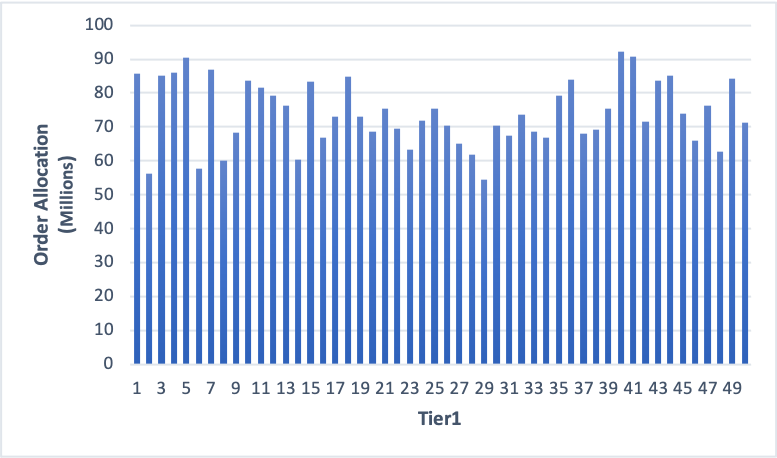}
				\label{subfig_allocation_machinists}
			\end{subfigure}%
			~ 
			\begin{subfigure}[t]{0.5\textwidth}			
				\centering
				\includegraphics[width=\textwidth]{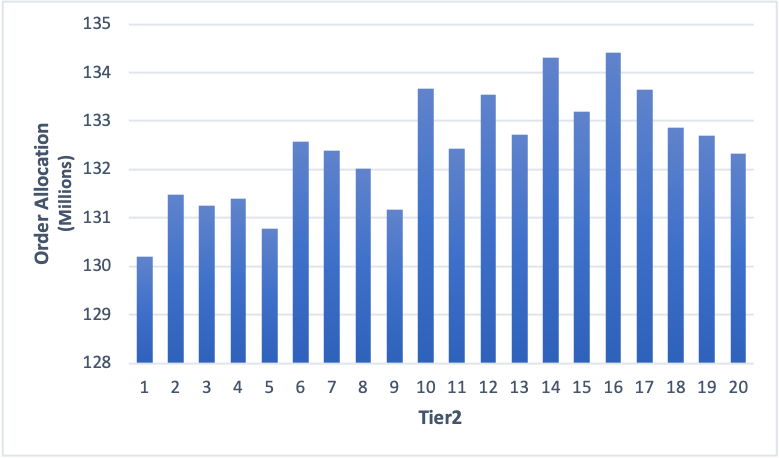}
				\label{subfig_allocation_forgers}
			\end{subfigure}
			\caption{Final allocations to suppliers.}
			\label{fig_allocations}
		\end{figure*}
	
		Final allocations of parts to Tier1 and allocations of forgings to Tier2 are are presented in Fig.~\ref{fig_allocations}. From the figure, we find that all the suppliers are allocated some orders which is required to keep the relationship active with the suppliers. Moreover, each supplier gets orders greater than the penalty threshold so the manufacturer is able to avoid penalty constraints (please refer to following subsection for sensitivity analysis using penalty factor and penalty threshold). This is mainly because of random data generation and keeping the threshold limit low.
        
        \paragraph{Comparative study of meta-heuristic algorithms:}
        
        Here, we compare the three most popular meta-heuristic algorithms, i.e., genetic algorithm (GA), particle swarm optimisation (PSO) and ant colony optimisation (ACO) for solving order allocation problems (\cite{Aouadni2019}). We compare the three algorithms for solving machinist-tier order allocations for single- as well as dual-sourcing. This is first due to the computational complexity of the SSOA probelm and second these algorithms do not improve the solutions provided by the exact optimisation method.
        
\begin{figure}[htb!]
            \centering
            \includegraphics[width=0.6\textwidth]{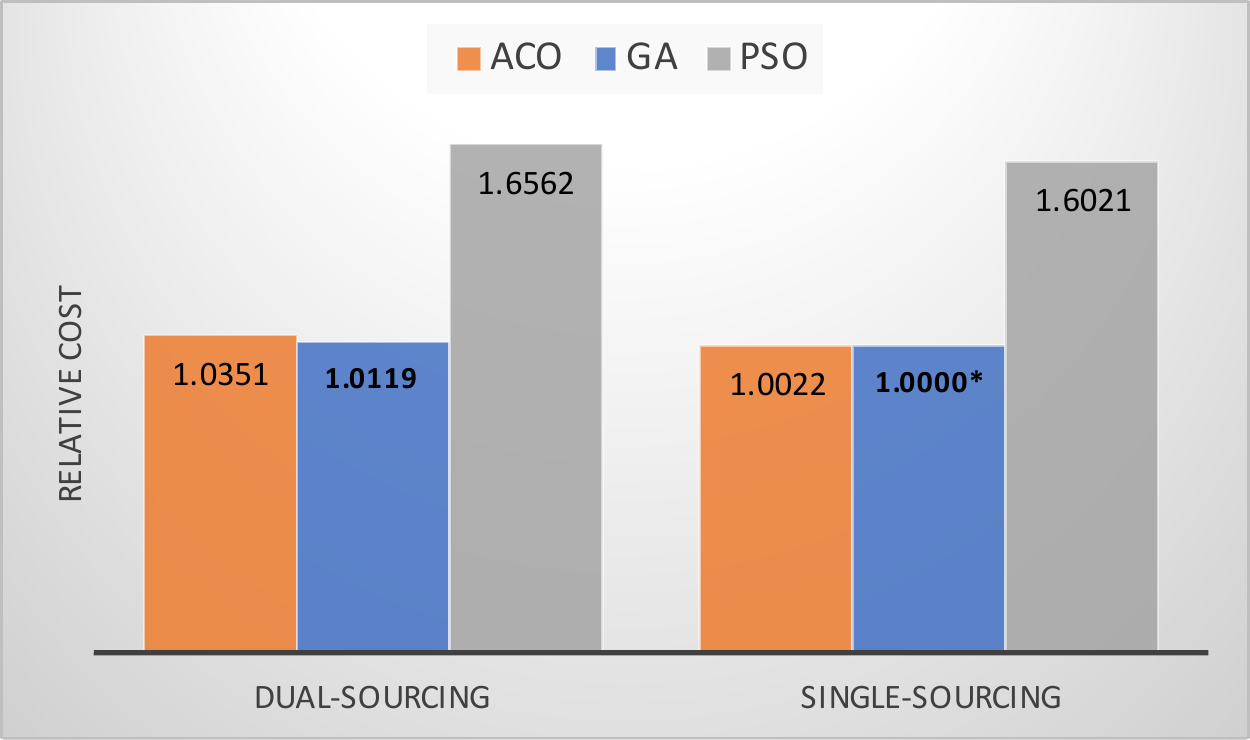}
            \caption{Comparative study of meta-heuristic algorithms for solving machinist-tier assignments. Y-axis represents the relative procurement cost, i.e., cost obtained from the algorithm is divided by the optimal solution (found using exact optimisation method). * represents the optimal solution.}
            \label{fig_metaheuristics}
        \end{figure}

        The comparative study of the three meta-heuristic algorithms for machinist-tier assignments is presented in the Fig.~\ref{fig_metaheuristics}, which compares the relative procurement cost, i.e., cost divided by the optimal cost obtained from the exact optimisation method. The dual-sourcing takes the case of 70:30 to calculate the procurement cost. From these results, we observe following points: (i) the GA algorithm outperforms the rest of the meta-heuristic algorithms in solving the SSOA problem for single- as well as dual-sourcing, and perhaps, that's why it is the most widely used meta-heuristic algorithm for order allocations (\cite{Aouadni2019}). (ii) Only the GA algorithm converges to the optimal solution, however, only for the single-sourcing problem. (iii) The ACO is the second best meta-heuristic algorithm and the PSO performs the worst for solving the SSOA problem in our study. The ACO performs very close to the the GA but there is large gap in performance for the PSO. (iv) It is also noted that all the three algorithms, relatively, converge to a better solution for single-sourcing than the dual-sourcing, which comparatively represents a simpler problem than the dual-sourcing.
        
        We have run all the three algorithms for sufficiently large values so that all of them get chance to converge. We observe that the GA takes 49 and 95 minutes, the PSO takes 158 and 303 minutes, and the ACO takes 1331 and 1821 minutes, for single- and dual-sourcing problems, respectively. Thus, the GA also converges faster than the rest of the algorithms, in addition to providing the best procurement cost. For algorithmic details and procedure of parameter selection, please refer to the Appendix~\ref{app_metaheuristic}.
    		
		\paragraph{Comparison with existing approach:}As described in Section~\ref{subsec_description}, earlier the allocations and negotiations were carried out manually by the experienced procurement managers which did not allow them sufficient time to negotiate and produce optimal allocations. In addition, the manufacturing company had also developed integer programming based tool\footnote{The tool was not shared with us as it was proprietary software.} which was taking more than 16 hours to solve the simplest case of machinist allocations and was not able to handle the complexity of forger and integrated allocations. That means the existing tool was impractical to solve SSOA and the all the allocations were performed manually. The proposed models helped the company to automate the manual allocations and produce optimal solutions (for machinist and forger allocations). The proposed models performed well as compared with models developed by the company potentially due to better modelling approach, as described in Section~\ref{subsec_mp}
		
		\subsection{Sensitivity analysis}
		\label{subsec_sensitivity}
\begin{figure*}[htb!]
			\centering
			\begin{subfigure}[t]{0.5\textwidth}			
				\centering
				\includegraphics[width=\textwidth]{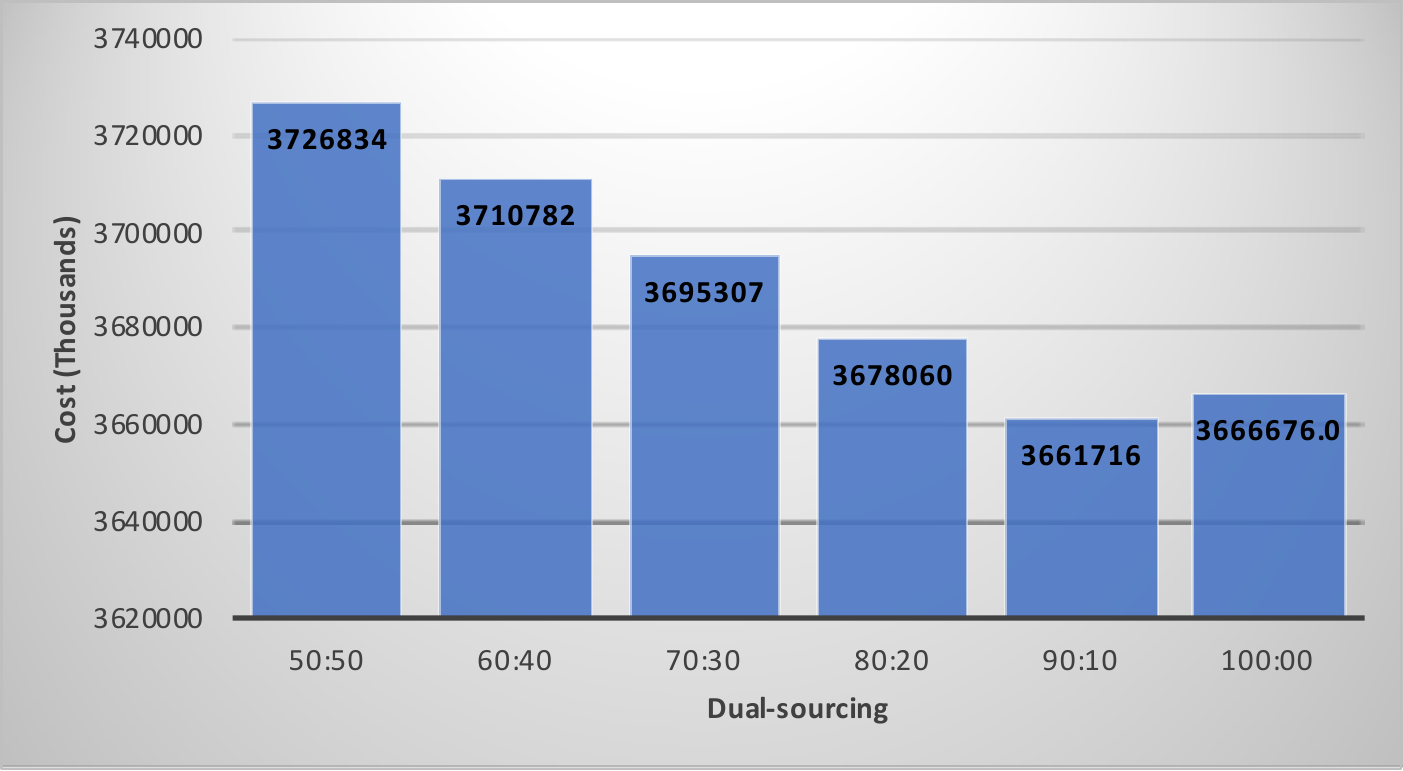}
				\caption{Machinist-tier}
				\label{subfig_machinistDS}
			\end{subfigure}%
			~ 
			\begin{subfigure}[t]{0.5\textwidth}			
				\centering
				\includegraphics[width=\textwidth]{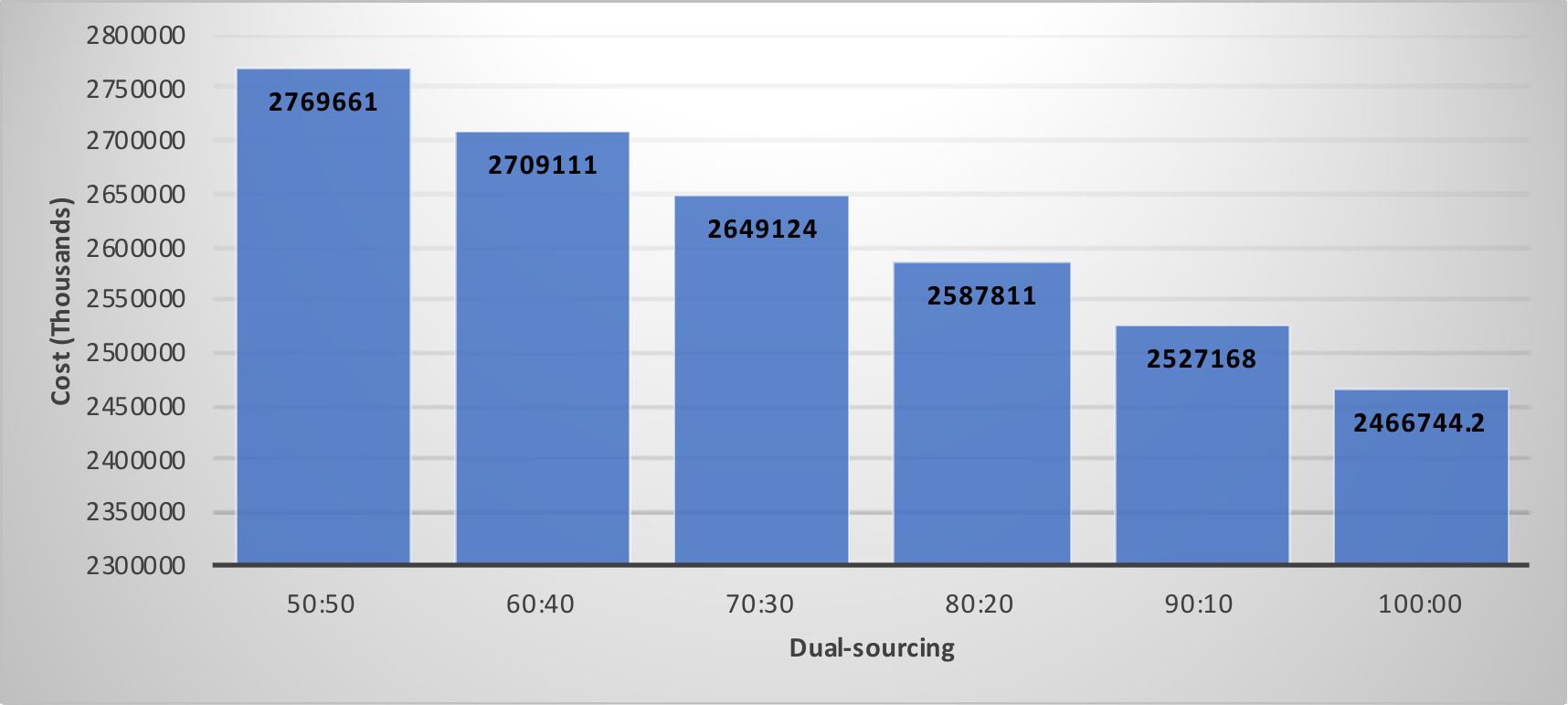}
				\caption{Forger-tier}
				\label{subfig_forgerDS}
			\end{subfigure}
			~ 
			\begin{subfigure}[t]{0.5\textwidth}			
				\centering
				\includegraphics[width=\textwidth]{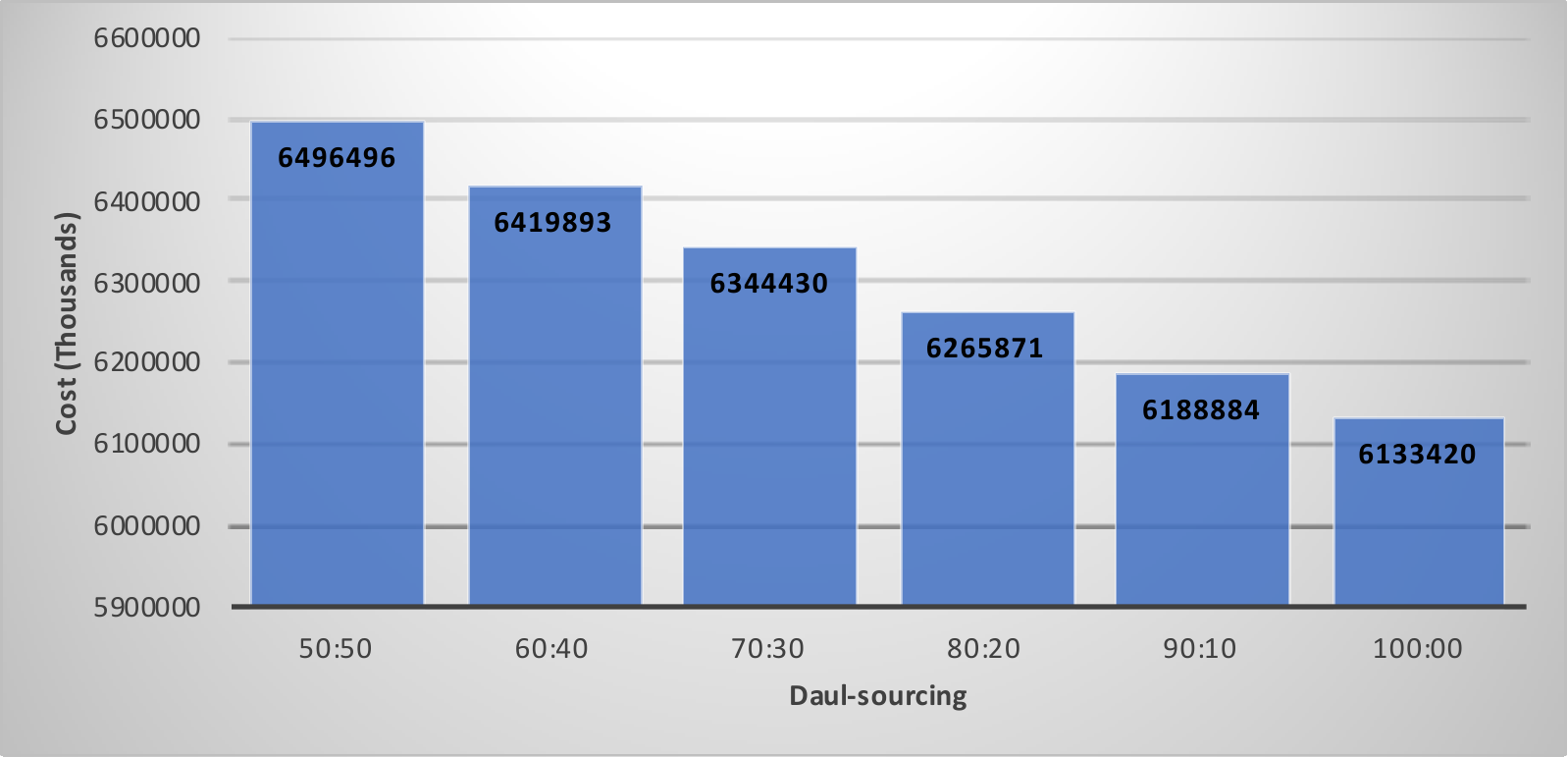}
				\caption{Integrated}
				\label{subfig_integratedDS}
			\end{subfigure}		
			\caption{Sensitivity analysis using sourcing-strategy for the three problems.}
			\label{fig_dualsourcing}
		\end{figure*}

		Fig.~\ref{fig_dualsourcing} represents sensitivity analysis of sourcing strategy for all the three problems, which
		shows interesting patterns. For dual-sourcing, as the difference between the two proportions increases, the procurement cost decreases, and becomes lowest for 100:00 (i.e., for the single-sourcing) for forger-tier and integrated problem, and at 90:10 for machinist-tier problem. In general, dual-sourcing is expected to be costlier than single-sourcing (\cite{Silbermayr2016}), because for dual-sourcing, assuming there is only one cheapest supplier, we are purchasing only some portion of the order from the cheapest supplier and the rest from another expensive supplier. On the other hand, single-sourcing cannot always be the cheapest because of the constraints that restrict us to order from the cheapest supplier, e.g., suppose buyer has a contract with a supplier to give some order for a part but that supplier is the most expensive. So, in single-sourcing the whole order for that part has to be given to expensive supplier but in dual-sourcing only some portion of order has to be given to the supplier. The two taken together, explains the best results for machinist-tier at 90:10. The company uses dual-sourcing because of two main reasons: (i) to avoid the monopoly of items by the supplier and, (ii) to reduce the risk of disruptions in the supply-chain as the company can continue to operate when one supplier is disrupted (\cite{Tomlin2006}).
		
		In practice, the company prefers to use 70:30 as dual-sourcing. This is because as we move towards the right of dual-sourcing spectrum, say at 90:10, one of the suppliers might not get sufficient quantities of items to justify its production expenses , making the solution infeasible in practice. On the other hand, moving to the left of the dual-sourcing spectrum, say at 50:50, the order assignments are expensive for the company, hence a suitable trade off between the two must be found.
		
\begin{figure*}[htb!]
			\centering
			\begin{subfigure}[t]{0.5\textwidth}			
				\centering
				\includegraphics[height=4cm]{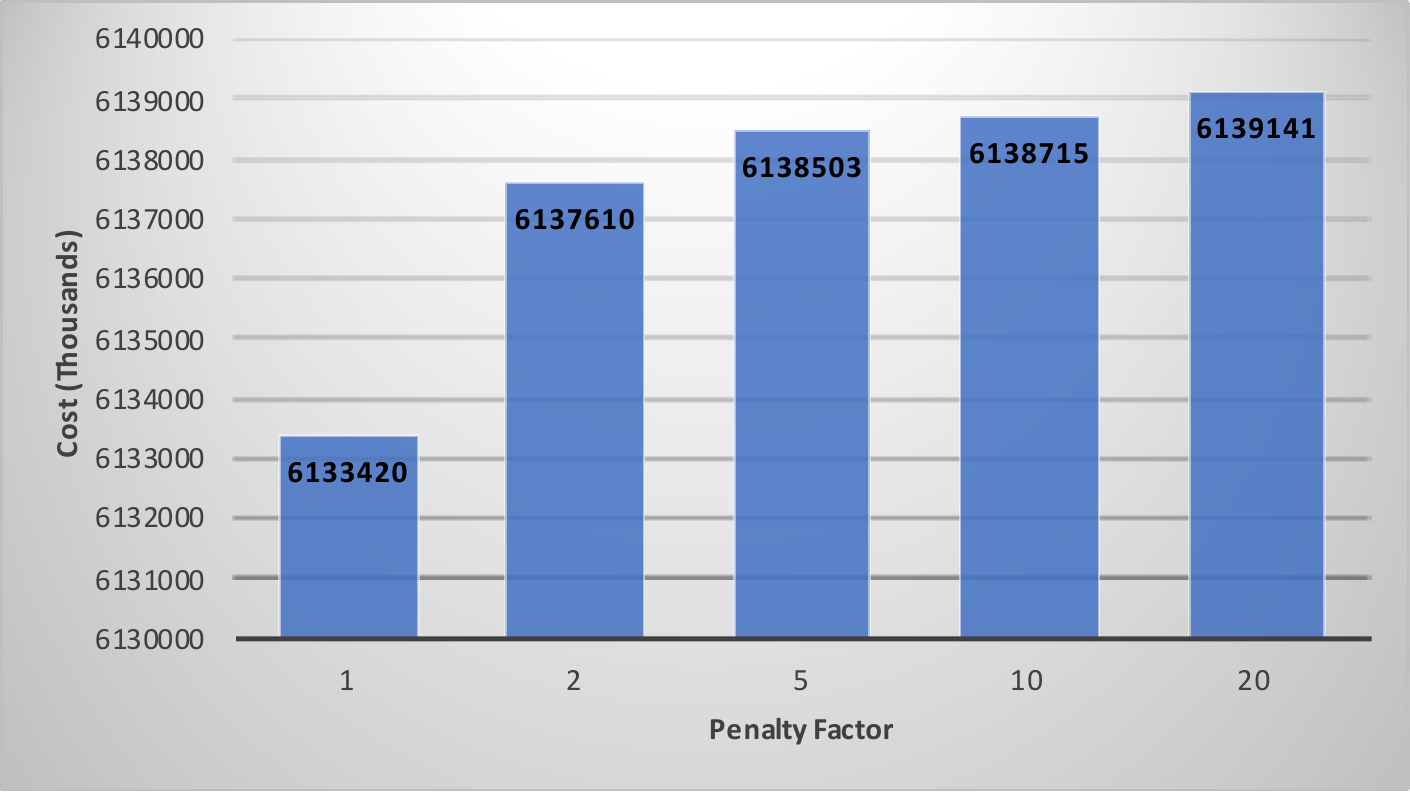}
				\caption{Effect of penalty factor on the procurement cost.}
				\label{subfig_penaltyFactor}
			\end{subfigure}%
			~ 
			\begin{subfigure}[t]{0.5\textwidth}			
				\centering
				\includegraphics[height=4cm]{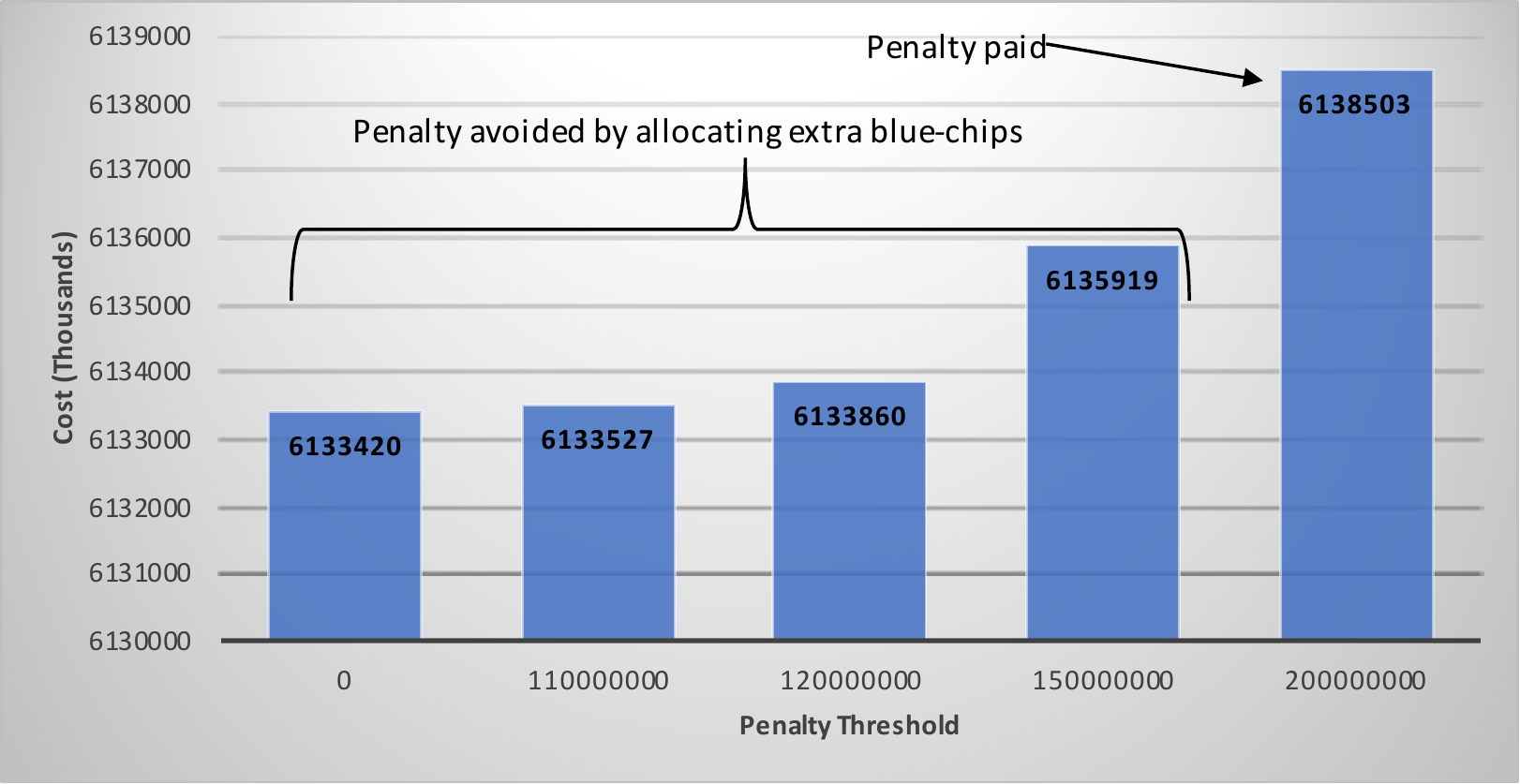}
				\caption{Effect of penalty threshold on the procurement cost.}
				\label{subfig_penaltyThreshold}
			\end{subfigure}	
			\caption{Sensitivity analysis of penalty factor and penalty threshold on integrated problem with single-sourcing.}
			\label{fig_penalty}
		\end{figure*}

		Fig.~\ref{fig_penalty} studies the effect of the penalty factor on the procurement cost using integrated problem with single-sourcing. For the sake of simplicity, the effect of penalty is studied only with one supplier. Subfig.~\ref{subfig_penaltyFactor} studies the effect of penalty factor at penalty threshold of 2.0E+8, and Subfig.~\ref{subfig_penaltyThreshold} studies the effect of penalty threshold with a penalty factor of five. From both of the sub-figures, it is clear that procurement cost increases with an associated increase in the penalty factor as well as the penalty threshold. Moreover, penalty constraints increase procurement cost because the buyer has to allocate a minimum amount, called penalty threshold of blue-chip orders, otherwise the supplier increases per unit cost of LLVs by a penalty factor. This results in two scenarios: first, where the buyer allocates sufficient blue-chip orders to the supplier to avoid the penalty at some added cost, and second, where the buyer does not allocate sufficient blue-chip orders to the supplier so the supplier increases the cost of LLVs resulting in increased procurement cost. Both the scenarios are highlighted by Subfig.~\ref{subfig_penaltyThreshold}, where the buyer avoids penalty for first four penalty threshold values but gets penalized at the threshold of 2.0E+8.
		
		The proposed model for forger-tier was successfully deployed by the company in a large international sourcing conference for a key engine programme for assignments of forgings by Tier1 to Tier2. The model was also used by the company to study what-if scenarios and to design the supply chain by studying the impact alternative configurations. The projected procurement cost of the conference was \$1bn and resulted in an estimated savings of more than 10\%.
		Since no mathematical models were used in earlier conferences the results could not be compared with other techniques. Moreover, the models are expected to be used in more sourcing conferences by the manufacturer as the manufacturer organizes five to six sourcing conferences every year.
		
		\section{Conclusion and discussions}
		\label{sec_conclusion}
		Supplier selection and order allocation are key strategic decisions in the supply chain management making the problem one of those that have been widely studied. However, there is a significant gap between theory and practice. Extant studies focus on variations of the problem where objectives and constraints are changed, however, little to no attention is given to large-scale case studies, particularly in the context of iterative sourcing auctions/conferences where suppliers bid on orders. In this context, the time complexity of the problem becomes important, which is mediated both by the problem scale, resulting formulation, and selection of suitable solution approaches. As a result of this discrepancy between theory and practice, practitioners are not well informed of the capabilities of optimization, and resort to using manual approaches resulting in sub-optimal solutions. 
		
		This paper aimed at addressing this gap by presenting a large-scale case study of supplier selection and order allocation problem across the two-tiers of a supply chain of a manufacturing company with supplier induced penalties and dual-sourcing. A novel problem was presented, where supplier selection and order allocation were performed across both the tiers of the supply chain with multi-to-multi relationships between items of the two tiers, real-time solution constraints, penalty based on item type and orders received by supplier, and supplier preferences to work with other suppliers through bidding.
		
		Mixed-Integer Programming models were developed for the individual-tiers and integrated problems. These problems are complex due to NP-hard nature, large-scale, real-time constraints, penalty constraints and multi-to-multi relationships between two tiers of the supply chain.
		
		The problems are solved using Mathematical Programming (MP) and Genetic Algorithms (GAs). The experimental results revealed that MP approach outperforms GA, both in terms of cost and time to solve individual tier and integrated tier problems. This is because GAs are memory and computation intensive and are suitable to solve complex problems where complexity is based on the non-linearity of problem and not on the scale of problem. However, finding a direct solution to the integrated problem is computationally infeasible due to scale and quadratic nature of the mixed-integer programming model so a 2-phase approach was developed, which decomposes the problem into two linear problems. The GA solutions are not suitable to the problem as it takes long time to solve the problem. Another interesting result is revealed by the sensitivity analysis of sourcing strategy. It is observed that the more balanced the dual-sourcing is the more expensive is the order assignment, and for extreme case of 100:00, i.e., single-sourcing the order assignments are cheapest for forger-tier and integrated problems.
		
		From a managerial perspective, this paper has developed models which facilitate procurement managers to take decisions in large scale sourcing problems such as supplier conferences to select suitable suppliers and allocate orders according to the bids submitted by the suppliers. The study also helps procurement managers to apply optimization to effectively negotiate prices through multiple real-time biddings. The study also helps managers to perform what-if analysis, i.e., removing certain suppliers and/or products from the supply chain or by shifting orders between suppliers, and studying their impact. This work helps in the digitalisation of complex supply chains by automating manual supplier selection processes. Moreover, double allocations between two-tiers of a supply chain also help improve cooperation and performance within the supply chain.
		
		Our approach has a number of limitations which present avenues for future extension. Trialling a larger number of meta-heuristic approaches and machine learning algorithms will help give better recommendations. Current work could also be extended by considering ways to incorporate more objectives such as to reduce the carbon emissions for an environmentally friendly supply chain, and disruption risks for building resilient supply chains. 
		
			\section*{Acknowledgement}
			The authors gratefully acknowledge the editor-in-chief, the associate editor and the anonymous reviewers for their constructive comments for improving the quality of the paper.
		

		\newpage
		\appendix 
		
		\section{linearization of integrated problem}
		\label{app_lin}
		This section discusses linearization of the integrated problem, and for simplicity, single-sourcing problem ($I_S$) is considered, which is given below.
		\begin{equation}
			\label{eq_obj_int_dual_}
			\begin{split}
				\min_{X, Y} \sum_{i_B,j} X_{i_B,j} \times CX_{i_B,j} + \sum_{i_L,j} X_{i_L,j} \times CX_{i_L,j}+ \sum_{k_B,j,l} Y_{k_B,j,l} \times CY_{k_B,j,l} + \sum_{k_L,j,l,d} Y_{k_L,j,l,d} \times CY_{k_L,j,l,d},
			\end{split}
		\end{equation}
		where, we have,
		\begin{equation}
			CY_{k_B,j,l} = \left(CBY_{k_B,j,l} + CTY_{k_B,j,l}\right) \times Z_{k_B,j},
		\end{equation}
		\begin{equation}
			CY_{k_L,j,l,d} = \left(CBY_{k_L,j,l} \times \gamma^d_l + CTY_{k_L,j,l}\right) \times Z_{k_L,j},
		\end{equation}
		\begin{equation}
			Z_{k_B,j} = \sum_{i_B}\left(X_{i_B,j} \times y_{i_B,k_B} \times NpiB \right) + \sum_{i_L}\left(X_{i_L,j} \times y_{i_L,k_B} \times NpiL \right),
		\end{equation}
		\begin{equation}
			\begin{split}
				Z_{k_L,j} = \sum_{i_B}\left(X_{i_B,j} \times y_{i_B,k_L} \times NpiB \right) + \sum_{i_L}\left(X_{i_L,j} \times y_{i_L,k_L} \times NpiL \right).
			\end{split}
		\end{equation}
		Last two terms of objective function (\ref{eq_obj_int_dual_}) are quadratic so let's simplify them by substituting the values of above equations,
		\begin{equation}
			\begin{aligned}
				\sum_{k_L,j,l,d} Y_{k_L,j,l,d} \times CY_{k_L,j,l,d} &\\
				=& \sum_{k_L,j,l,d} Y_{k_L,j,l,d} \times \left(CBY_{k_L,j,l} \times \gamma^d_l + CTY_{k_L,j,l}\right) \times Z_{k_L,j}\\
				=& \sum_{k_L,j,l,d} Y_{k_L,j,l,d} \times \left(CBY_{k_L,j,l} \times \gamma^d_l + CTY_{k_L,j,l}\right) \times \biggl(\sum_{i_B}X_{i_B,j} \times y_{i_B,k_L} \times NpiB \\ 
				&+ \sum_{i_L}X_{i_L,j} \times y_{i_L,k_L} \times NpiL \biggr),\\
				=& \sum_{k_L,j,l,d,i_B,i_L} Y_{k_L,j,l,d} \times \left(CBY_{k_L,j,l} \times \gamma^d_l + CTY_{k_L,j,l}\right) \times \biggl(X_{i_B,j} \times y_{i_B,k_L} \times NpiB \\ 
				&+ X_{i_L,j} \times y_{i_L,k_L} \times NpiL \biggr),\\
				=& \sum_{k_L,j,l,d,i_B,i_L} \left(CBY_{k_L,j,l} \times \gamma^d_l + CTY_{k_L,j,l}\right) \times \biggl(Y_{k_L,j,l,d} \times X_{i_B,j} \times y_{i_B,k_L} \times NpiB \\ 
				&+ Y_{k_L,j,l,d} \times X_{i_L,j} \times y_{i_L,k_L} \times NpiL \biggr).
			\end{aligned}
		\end{equation}
		Now, let's introduce following new binary variables and substituting in previous equation, we get,
		\begin{equation}
			\begin{split}
				U_{k_L,j,l,d,i_B} = Y_{k_L,j,l,d} \times X_{i_B,j},\\
				U_{k_L,j,l,d,i_L} = Y_{k_L,j,l,d} \times X_{i_L,j}.
			\end{split}
		\end{equation}
		\begin{equation}
			\begin{aligned}
				\sum_{k_L,j,l,d} Y_{k_L,j,l,d} \times CY_{k_L,j,l,d} &\\
				=& \sum_{k_L,j,l,d,i_B,i_L} \left(CBY_{k_L,j,l} \times \gamma^d_l + CTY_{k_L,j,l}\right) \times \biggl(U_{k_L,j,l,d,i_B} \times y_{i_B,k_L} \times NpiB \\ 
				&+ U_{k_L,j,l,d,i_L} \times y_{i_L,k_L} \times NpiL \biggr),
			\end{aligned}
		\end{equation}
		Similarly, we can simplify the other term, by introducing new variables, as given below.
		\begin{equation}
			\begin{aligned}
				\sum_{k_B,j,l} Y_{k_B,j,l} \times CY_{k_B,j,l} &\\
				=& \sum_{k_B,j,l,i_B,i_L} \left(CBY_{k_B,j,l} + CTY_{k_B,j,l}\right) \times \biggl(U_{k_B,j,l,i_B} \times y_{i_B,k_B} \times NpiB \\ 
				&+ U_{k_B,j,l,i_L} \times y_{i_L,k_B} \times NpiL \biggr),
			\end{aligned}
		\end{equation}
		where,
		\begin{equation}
			\begin{split}
				U_{k_B,j,l,i_B} = Y_{k_B,j,l} \times X_{i_B,j},\\
				U_{k_B,j,l,i_L} = Y_{k_B,j,l} \times X_{i_L,j}.
			\end{split}
		\end{equation}
		Now, substituting the above simplified terms into the objective function, we get a linear mixed-integer programming problem. The linearized problem has following new constraints due to linearization, in addition to constraints of the original non-linear problem. The forger-budget constraint, i.e., inequality (\ref{eq_F_S_budget}) is a quadratic constraint in the original problem which can be converted to linear by substituting the values of above simplified terms. The new constraints are as per the linearization rule discussed in \ref{subsubsec_formulation}, and for $U_{k_L,j,l,d,i_B} = Y_{k_L,j,l,d} \times X_{i_B,j}$, constraints can be defined as:
		\begin{equation}
			\begin{split}
				U_{k_L,j,l,d,i_B} \le X_{i_B,j}, \qquad U_{k_L,j,l,d,i_B} \le Y_{k_L,j,l,d},\\
				U_{k_L,j,l,d,i_B} \ge X_{i_B,j} + Y_{k_L,j,l,d} -1, \quad \forall k_L, j, l, d, i_B.
			\end{split}
		\end{equation}
		Similarly, we can define new constraints for remaining three new variables, i.e., $U_{k_L,j,l,d,i_L}, U_{k_B,j,l,i_B}$ and $U_{k_B,j,l,i_L}$. Moreover, integrated dual-sourcing problem ($I_D$) can be linearized on the similar lines.

		\section{Order assignments with single-sourcing}
		\label{app_single}
		Models with single sourcing can be easily obtained from dual-sourcing models by removing dual-sourcing
		constraints and equivalent notations can be obtained by simply removing superscripts denoting dual-sourcing proportions. Thus, corresponding costs can be calculated similar to dual-sourcing problems, as given below.
		\begin{equation}
			CX_{i_B,j} = \left(CBX_{i_B,j} + CTX_{i_B,j}\right) \times Npi_B,
		\end{equation}
		\begin{equation}
			CX_{i_L,j} = \left(CBX_{i_L,j} + CTX_{i_L,j}\right) \times Npi_L,
		\end{equation}
		\begin{equation}
			CY_{k_B,j,l} = \left(CBY_{k_B,j,l} + CTY_{k_B,j,l}\right) \times Z_{k_B,j},
		\end{equation}
		\begin{equation}
			CY_{k_L,j,l,d} = \left(CBY_{k_L,j,l} \times \gamma^d_l + CTY_{k_L,j,l}\right) \times Z_{k_L,j},
		\end{equation}
		where $Z_{k_B,j}$ and $Z_{k_L,j}$ are number of forgings $k_B$ and $k_L$ required by Tier1 supplier, respectively, as calculated below. 
		\begin{equation}
			\begin{split}
				Z_{k_B,j} = \sum_{i_B} \left(X_{i_B,j} \times y_{i_B,k_B} \times NpiB \right) + \sum_{i_L} \left(X_{i_L,j} \times y_{i_L,k_B} \times NpiL \right),
			\end{split}
		\end{equation}
		\begin{equation}
			\begin{split}
				Z_{k_L,j} = \sum_{i_B}\left(X_{i_B,j} \times y_{i_B,k_L} \times NpiB \right)+ \sum_{i_L}\left(X_{i_L,j} \times y_{i_L,k_L} \times NpiL \right). 
			\end{split}
		\end{equation}
		
		\paragraph{\textbf{Machinist-tier problem with single-sourcing ($M_S$)}:} This problem involves selecting Tier1 suppliers to supply finished parts and allocating quantities of parts with single sourcing
		It considers only machinist-tier and optimizes the machining cost for finished parts, as given below. 
		\begin{equation}
			\label{eq_M_S}
			\min _{X} \quad \sum_{i_B,j} X_{i_B,j} \times CX_{i_B,j} + \sum_{i_L,j} X_{i_L,j} \times CX_{i_L,j} ,
		\end{equation}
		subject to,
		\begin{equation}
			\label{eq_M_S_budget}
			S_{mj}^{min} \le \sum_{i_B,i_L} \left( X_{i_B,j} \times CX_{i_B,j} + X_{i_L,j} \times CX_{i_L,j} \right)  \le S_{mj}^{max}, \forall j,
		\end{equation}
		\begin{equation}
			\label{eq_M_S_mustB}
			X_{i_B,j} = 1, \qquad \text{for set of values of } \lbrace \left(i_B,j\right) \rbrace,
		\end{equation}
		\begin{equation}
			\label{eq_M_S_mustL}
			X_{i_L,j} = 1, \qquad \text{for set of values of } \lbrace \left(i_L,j\right) \rbrace.
		\end{equation}
		
		Here, equation (\ref{eq_M_S}) is the objective function and it calculates the machining cost of finished parts to the company as per the requirements. Inequality (\ref{eq_M_S_budget}) forces budget constraints and equalities (\ref{eq_M_S_mustB}) and (\ref{eq_M_D_mustL}) ensure must-make constraints on Tier1 for parts $i_B$ and $i_L$. $M_S$ is a mixed-integer linear programming problem.
		
		\paragraph{\textbf{Forger-tier problem with single-sourcing ($F_S$)}:} This problem involves selecting Tier2 to supply forgings to Tier1 and allocating quantities of forgings with single sourcing. It considers only forger-tier and optimizes the procurement cost for forgings by Tier1, as given below.
		\begin{equation}
			\label{eq_F_S}
			\min_{Y}  \sum_{k_B,j,l} Y_{k_B,j,l} \times CY_{k_B,j,l}  + \sum_{k_L,j,l,d} Y_{k_L,j,l,d} \times CY_{k_L,j,l,d} 
		\end{equation}
		\begin{equation}
			\label{eq_F_S_budget}
			S_{fl}^{min} \le \sum_{k_B,j} \left(Y_{k_B,j,l} \times CY_{k_B,j,l}\right) + \sum_{k_L,j,d} \left(Y_{k_L,j,l,d} \times CY_{k_L,j,l,d}\right) \le S_{fl}^{max} \quad \forall l
		\end{equation}
		\begin{equation}
			\label{eq_F_S_mustB}
			Y_{k_B,j,l} = 1, \quad \text{for set of values of } \lbrace \left(k_B,j,l\right) \rbrace
		\end{equation}
		\begin{equation}
			\label{eq_F_S_mustL}
			\sum_{d} Y_{k_L,j,l,d}= 1, \qquad \text{for set of values of } \lbrace \left(k_L,j,l\right) \rbrace
		\end{equation}
		\begin{equation}
			\label{eq_F_S_penalty1}
			-M v_l \le \sum_{k_B,j} \left(Y_{k_B,j,l} \times CY_{k_B,j,l} \right) -D_l +\epsilon \le M(1-v_l), \quad \forall l
		\end{equation}
		\begin{equation}
			\label{eq_F_S_penalty2}
			\sum_{k_L,j} \left(Y_{k_L,j,l,d_1} \times CY_{k_L,j,l,d_1}\right) \leq M(1-v_l), \quad \forall l
		\end{equation}
		\begin{equation}
			\label{eq_F_S_penalty3}
			\sum_{k_L,j} \left(Y_{k_L,j,l,d_2} \times CY_{k_L,j,l,d_2} \right) \leq Mv_l. \quad \forall l
		\end{equation}
		Here, equation (\ref{eq_F_S}) is the objective function and it calculates the cost of supplying forgings to Tier1 as per the requirements. Inequality (\ref{eq_F_S_budget}) forces budget constraints and equalities (\ref{eq_F_S_mustB}) and (\ref{eq_F_S_mustL}) ensure must-make constraints on Tier2 for forgings $k_B$ and $k_L$. Inequalities (\ref{eq_F_S_penalty1}) to (\ref{eq_F_S_penalty3}) are used to ensure penalty constraints. $F_S$ is also a mixed-integer linear programming problem.
		
		\paragraph{\textbf{Integrated problem with single-sourcing ($I_S$)}:} This problem considers both machinist and forger tiers and optimizes the overall cost of supply chain for selecting Tier1 and order allocation for parts and selecting Tier2 and order allocation for forgings. 
		\begin{equation}
			\label{eq_obj_int_dual}
			\begin{split}
				\min_{X, Y} \sum_{i_B,j} X_{i_B,j} \times CX_{i_B,j} + \sum_{i_L,j} X_{i_L,j} \times CX_{i_L,j}+ \sum_{k_B,j,l} Y_{k_B,j,l} \times CY_{k_B,j,l} + \sum_{k_L,j,l,d} Y_{k_L,j,l,d} \times CY_{k_L,j,l,d}.
			\end{split}
		\end{equation}
		
		The constraints on integrated problem are combined from constraints on machinist-tier and forger-tier problem. Inequalities (\ref{eq_M_S_budget}) and (\ref{eq_F_S_budget}) force upper and lower budget constraints on Tier1 and Tier2, respectively, and equalities (\ref{eq_M_S_mustB}), (\ref{eq_M_S_mustL}), (\ref{eq_F_S_mustB}) and (\ref{eq_F_S_mustL}) are used to ensure must-make constraints on Tier1 and Tier2, respectively. Inequalities (\ref{eq_F_S_penalty1}) to (\ref{eq_F_S_penalty3}) are used to ensure penalty constraints. $I_S$ is a mixed-integer quadratic linear programming problem with quadratic constraints.
		
		\section{Results with single-sourcing}
		\label{app_singlesourcing_results}
		Table~\ref{tab_mp_vs_ga_single} compares solutions obtained by MP and GA approaches for solving order assignment with single-sourcing in terms of time to solve the problem (in minutes) and ratio of procurement cost and best procurement cost obtained.
		\begin{table}[htb!]
			\centering
			\caption{MP versus GA with single-sourcing}
			\label{tab_mp_vs_ga_single}
			\begin{tabular}{|c|r|r|r|r|}
				\hline
				\multirow{2}{*}{\textbf{Problem}} & \multicolumn{2}{c|}{\textbf{MP}} & \multicolumn{2}{c|}{\textbf{GA}} \\ \cline{2-5} 
				& \multicolumn{1}{c|}{Time (minutes)} & \multicolumn{1}{c|}{Cost/best} & \multicolumn{1}{c|}{Time (minutes)} & \multicolumn{1}{c|}{Cost/best} \\ \hline
				\textit{machinist-tier} & 0.072 & 1.00000 & 49.040 & 1.00000 \\ \hline
				\textit{forger-tier} & 10.396 & 1.00000 & 3820.131 & 1.77228 \\ \hline
				\textit{Integrated} & 10.602 & 1.00000 & 5324.024 & 1.35460 \\ \hline
			\end{tabular}
		\end{table}
		From the table, it is clear that MP approach is faster than GA approach to solve all the three problems. Moreover, GA approach is not practical as it cannot solve the forger-tier and integrated problems in real-time which was a key requirement. MP also produces better solutions than GA, except the machinist-tier where both converge to optimal solution.

\section{Comparative study of meta-heuristic techniques}
\label{app_metaheuristic}
In this section, we provide the algorithmic details for PSO and ACO, and the automated selection of the best parameters for the both the algorithms.

\subsection{Particle swarm optimisation}
Particle swarm optimisation (PSO) is a biologically inspired meta-heuristic algorithm for solving optimisation problems, which is based on the idea that a flock of birds flying in a group ``can profit from the experience of all other members" (\cite{Kennedy1995}). We start with a swarm of random solutions each of which evolve by utilising it's own best position as well as the best solution of the whole swarm from one iteration to another.

To solve the SSOA problem using the PSO, we model the solution of SSOA similar to the GA solution, i.e., the solution is represented as a vector $x$ of length equal to number of parts which stores the suppliers supplying them (for dual-sourcing, the length of the vector is doubled to store two suppliers for each part). We implemented the following PSO algorithm to solve the SSOA (for details refer to \cite{poli2007particle}) and the notations used in the algorithm are defined in Table~\ref{tab_notations_pso}.

\begin{table}[htb!]
\centering
\caption{Notations for PSO algorithm}
\label{tab_notations_pso}
    \begin{tabular}{ll}\\ \hline
    \textbf{Variable} & \textbf{Definition} \\ \hline
    $k$ & iteration index \\
    $N$ & Number of particles in the swarm \\
    $x_k^i$ & position of particle $i$ in iteration $k$ \\
    $f_k^i$ & cost function at particle position $x_k^i$ \\
    $v_k^i$ & velocity of particle $i$ in iteration $k$ \\
    $p_k^i$ & best individual position of particle $i$ in iteration $k$ \\
    $p_k^g$ & best swarm position in iteration $k$ \\
    $w_k$   & inertia weight in iteration $k$\\
    $c_1, c_2$ & cognitive and social parameters, respectively \\
    $r_1, r_2$ & random numbers between 0 and 1 \\ \hline
    \end{tabular}
\end{table}

\noindent Step 1. Initialisation: Initialise the constants: $k_{max}, w_k, c_1, c_2, N$; generate $N$ particles and randomly assign positions and velocities to each of them.\\
Step 2. Repeat the algorithm for $k=1,2,...,k_{max}$ iterations.\\
Step i. For each particle $i$, enforce the constraints and evaluate the cost function $f_k^i$. We enforce the constraints by randomly changing the assignments which violate the constraints.\\
Step ii. For each particle $i$, save the best cost value and the best position/solution:
\begin{equation}
    \label{eq_update_best_i}
    \text{If  } f_k^i < f_{best}^i, \text{  then  } f_{best}^i = f_k^i \text{  and  } p_k^i = x_k^i.
\end{equation}
Step iii. For each particle $i$, compare with the swarm's best cost and save the best cost value and the best position/solution for the swarm:
\begin{equation}
    \label{eq_update_best_g}
    \text{If  } f_k^i < f_{best}^g, \text{  then  } f_{best}^g = f_k^i \text{  and  } p_k^g = x_k^i.
\end{equation}
Step iv. Update the position $x_k^i$ and velocity $v_k^i$ for each particle $i$ as given below.
\begin{equation}
    \label{eq_pso_xik}
    \begin{split}
    v_{k+1}^i =& w_k v_k^i + c_1 r_1 (p_k^i - x_k^i) + c_2 r_2 (p_k^g - x_k^i),\\
    x_{k+1}^i =& x_k^i + v_{k+1}^i.
    \end{split}
\end{equation}
Since the solution is an array of integers so we round each value of the position vector $x_k^i$ to the nearest integer. Each particle is under three forces: particle velocity, i.e., inertia ($w_k v_k^i$), best known position of the particle ($c_1 r_1 (p_k^i - x_k^i)$) and the best position of the swarm ($c_2 r_2 (p_k^g - x_k^i)$). These three factors are controlled with constants $w_k, c_1, c_2$ and random numbers $r_1, r_2$.\\
Step 3. Return the swarm's best solution.

\subsection{Ant colony optimisation}
Ant colony optimisation (ACO) is another meta-heuristic swarm optimisation algorithm, which is motivated by the movement of the ants between their colony and the food source (\cite{dorigo1991positive}). As ants go out in search for food and return to their colony, they leave behind pheromone on their way which helps them to trace their path. So as more and more ants follow a path, the more is the pheromone on that path and higher is the probability of the ants to follow that path in comparison to several other paths followed by different ants. As discussed below, the ACO algorithm uses the same approach and stores pheromone for each edge on the graph, and the probability of selecting that edge in the solution is calculated based on the pheromones.

To solve the SSOA problem using the ACO, we model the SSOA as a bi-partite graph, where one set of nodes refers to the suppliers and the other set of nodes refers to the parts to be supplied by the suppliers. The solution of SSOA consists of edges from parts to suppliers -- referring to assignments. So, the solution must contain an edge from each part to suppliers. We implemented the following ACO algorithm to solve the SSOA (for details refer to \cite{Dorigo2006}) and the notations used in the algorithm are defined in Table~\ref{tab_notations_aco}.
\begin{table}[htb!]
\centering
\caption{Notations for ACO algorithm}
\label{tab_notations_aco}
    \begin{tabular}{ll}\\ \hline
    \textbf{Variable} & \textbf{Definition} \\ \hline
    $k=1,2,..,N$ & index for ant and $N$ is max number of ants \\
    $\tau_{ij}$ & pheromone for supplying of part $i$ by supplier $j$\\
    $\nabla \tau_{ij}^k$ & pheromone collected in one iteration for supplying of part $i$ by supplier $j$ by ant $k$\\
    $p_{ij}$ & probability of supplying part $i$ by supplier $j$ \\
    $d_{ij}$ & cost of supplying part $i$ by supplier $j$ \\
    $\eta_{ij}$ & visibility and calculated as $1/d_{ij}$\\
    $\rho$ & evaporation constant \\
    $\alpha, \beta$ &  constants to control the strength of pheromone and visibility\\
    $Q$ &  a constant used in calculating pheromone\\
    $L_k$ & cost of solution $k$, i.e., cost of the tour of the ant $k$ \\
    $max\_iter$ & max number of iterations to run for solution \\ \hline
    \end{tabular}
\end{table}

\noindent Step 1. Initialisation: Initialise constants $\alpha, \beta, Q, \rho$, max\_iter and $\tau_{ij}$ randomly to small values, create $N$ ants, and also calculate the initial probabilities of assigning parts to suppliers, as given below.
\begin{equation}
    \label{eq_aco_prob}
    p_{ij} = \dfrac{\tau_{ij}^\alpha \eta_{ij}^\beta}{\sum_{j} \tau_{ij}^\alpha \eta_{ij}^\beta},
\end{equation}
where $\eta_{ij} = 1/d_{ij}$ is the visibility and $\tau$ and $\eta$ trade-off global and local factors in finding the assignments.\\
Step 2. Run the following steps for max\_iter:\\
Step i. Generate a tour, i.e., solution for each ant $k$ as per the probabilities $p_{ij}$\\
Step ii. Calculate the cost $L_k$ of the solution for each ant $k$, while enforcing constraints by randomly changing the violating assignments.\\
Step iii. Evaporate pheromone as given below.
\begin{equation}
    \tau_{ij} = (1 - \rho) \tau_{ij},
\end{equation}
where $0 < \rho < 1$ is an evaporation constant.\\
Step iv. Update the pheromone as per the new assignments found by the ants, as given below.
\begin{equation}
    \tau_{ij} = \tau_{ij} + \sum_k \nabla \tau_{ij}^k,
\end{equation}
where $\nabla \tau_{ij}^k = Q / L_k$.\\
Step v. Update the probabilities $p_{ij}$ of assigning parts $i$ to suppliers $j$ from the updated pheromone calculations using the Eq.~(\ref{eq_aco_prob}).\\
Step 3. Return the best ant tour.

\subsection{Parameter tuning}
\begin{table}[htb!]
\centering
\caption{Parameters selection for ACO and PSO algorithms - final parameters are selected through automated random search using HyperOpt library.}
\label{tab_hyperopt}
    \begin{tabular}{lcr}\\ \hline
    \textbf{Parameter} & \textbf{Range} & \textbf{Selected Value} \\ \hline
    ACO & & \\ \hline
    N & $\left[20, 120\right]$ & 100 \\
    $\alpha$ & $\left[0.5, 1.5\right]$ & 1.0 \\
    $\beta$ & $\left[0.5, 2.0\right]$ & 1.5 \\
    $\rho$ & $\left[0.1, 0.4\right]$ & 0.2 \\ \hline
    PSO & & \\ \hline
    N & $\left[50, 200\right]$ & 200\\
    $w_k$ & $\left[0.3, 0.8\right]$ & 0.4 \\
    $c_1$ &$\left[1, 4\right]$ & 3 \\
    $c_2$ &$\left[1, 4\right]$ & 1 \\ \hline
    \end{tabular}
\end{table}
PSO as well as ACO are sensitive to the choice of parameters so to find the best parameter values, we have used automated search using the library HyperOpt\footnote{http://hyperopt.github.io/hyperopt/} (\cite{bergstra2013making}), which applies random search. The final selected parameters and the range of values searched are provided in the Table~\ref{tab_hyperopt}. For ACO, the value of $Q$ is selected 10000000.0 to ensure the convergence and then the rest of the values are searched because the algorithm converges only if right value is selected. Moreover, the ACO is run for 5000 iterations as it does not show further improvement and it's each iteration is computationally very expensive. The initial values of $\tau_{ij}$ are set small in $\left[0, 1e-3\right]$. The PSO is run for 50000 iterations. The experiments are averaged over five runs starting with different random seeds.

\end{document}